\documentclass{aa}

\usepackage{graphicx}     
\usepackage{pdflscape} 
\usepackage{color}
\usepackage{natbib}
\usepackage{mathrsfs}
\usepackage{fontenc} 
\usepackage{longtable}
\usepackage{txfonts} 

\begin{document} 

\title{Be and Bn stars: Balmer discontinuity and stellar-class relationship\thanks{Based on observations obtained at the Complejo Astron\'omico El Leoncito (CASLEO), operated under an agreement between the Consejo Nacional de Investigaciones Cient\'{\i}ficas y T\'ecnicas de la Rep\'ublica Argentina, the Secretar\'{\i}a de Ciencia y Tecnolog\'{\i}a de la Naci\'on and the National Universities of La Plata, C\'ordoba and San Juan, Argentina}}

\author{Y.~R. Cochetti\inst{1}
\and    J. Zorec\inst{2}
\and    L.~S. Cidale\inst{1,3}
\and    M.~L. Arias\inst{1,3}
\and    Y. Aidelman\inst{1}
\and    A.~F. Torres\inst{1,3}
\and    Y. Fr\'emat\inst{2,4}
\and    A. Granada\inst{3,5}
}
\institute{Departamento de Espectroscopía Estelar, Facultad de Ciencias Astronómicas y Geofísicas, Universidad Nacional de La Plata, La Plata, Argentina\\ \email{cochetti@fcaglp.unlp.edu.ar}
\and Sorbonne Universit\'e, CNRS, UPMC, UMR 7095, Institut d'Astrophysique de Paris, 98 bis bd. Arago, 75014 Paris, France\\ \email{zorec@iap.fr}
\and Instituto de Astrof\'isica de La Plata, CCT-La Plata, CONICET-UNLP, Paseo del Bosque s/n, CP 1900, La Plata, Buenos Aires, Argentina
\and Royal Observatory of Belgium, 3 Av. Circulaire, B-1180 Bruxelles, Belgium
\and Laboratorio de Procesamiento de Se\~nales Aplicadas y Computación de Alto Rendimiento, Sede Andina, Universidad Nacional de Rıo Negro, Mitre 630, San Carlos de Bariloche, R8400AHN R\'io Negro, Argentina
}

\date{Received , 2019; accepted , 2019}

\abstract
{A significant number of Be stars show a second Balmer discontinuity (sBD) attributed to an extended circumstellar envelope (CE). The fast rotational velocity of Be stars undoubtedly plays a significant role in the formation of the CE. However, Bn stars, which are also B-type rapidly rotating stars, do not all present clear evidence of being surrounded by circumstellar material.}
{We aim to characterize the populations of Be and Bn stars, and discuss the appearance of the sBD as a function of the stellar parameters. We expect to find new indices characterizing the properties of CEs in Be stars and properties relating Be and Bn stars.} 
{We obtained low- and high-resolution spectra of a sample of Be and Bn stars, derived stellar parameters, characterized the sBD, and measured the emission in the H$\alpha$ line.}
{Correlations of the aspect and intensity of the sBD and the emission in the H$\alpha$ line with the stellar parameters and the $V\!\sin i$ are presented. Some Bn stars exhibit the sBD in absorption, which may indicate the presence of rather dense CEs. Six Bn stars show emission in the H$\alpha$ line, so they are reclassified as Be stars. The sBD in emission appears in Be stars with $V\!\sin i \lesssim 250$ km\,s$^{-1}$, and in absorption in both Be and Bn stars with \mbox{$V\!\sin i \gtrsim 50$ km\,s$^{-1}$}. Low-mass Be and Bn stars share the same region in the Hertzsprung-Russell diagram. The distributions of rotational to critical velocity ratios of Be and Bn stars corresponding to the current stellar evolutionary stage are similar, while distributions inferred for the zero-age main sequence have different skewness.}
{We found emission in the H$\alpha$ line and signs of a CE in some Bn stars, which motivated us to think that Bn and Be stars probably belong to the same population. It should be noted that some of the most massive Bn stars could display the Be phenomenon at any time. The similarities found among Be and Bn stars deserve to be more deeply pursued.}
 

\maketitle

\section{Introduction}\label{intr} 

Be stars are non-supergiant B-type stars that are    rapidly rotating and surrounded by a gaseous extended circumstellar envelope (CE) whose structure and formation physics are still under debate. This envelope gives rise to a wide variety of spectroscopic peculiarities that characterize the Be phenomenon \citep[see][]{Porter2003,Rivinius2013}. Particularly, in the optical range Be stars show, or have shown at least once, hydrogen lines in emission \citep{Jaschek1981}. Their continuum spectra exhibit flux excesses mainly in the optical and near-infrared regions and, in several cases, they exhibit two components of the Balmer discontinuity \citep[hereafter BD;][]{BarbierChalonge1939,ChalongeDivan1952,Schild1978,Divan1979}. The first component of the BD, $D^*$, is stellar photospheric. It is constant and defines the spectral type of the stellar hemisphere projected toward the observer, as it does for emissionless non-supergiant B-type stars. The second component of the BD (sBD, $d$) appears at shorter wavelengths, very close to the theoretical Balmer line series limit. This second component of the BD originates in a low pressure stellar gaseous environment. It is variable and can be either in emission or in absorption. At times, it can completely disappear \citep[e.g.,][]{Divan1979,Underhill1982,Zorec1986,ZorecBriot1991,Moujtahid1998,Aidelman2012}. \par

\citet{Schild1978}, \citet{Kaiser1987,Kaiser1989} and \citet{Dachs1989} presented a series of papers with measurements of $d$ for a reduced number of northern and southern non-shell Be stars. These authors found trends or vague correlations between $d$ and the emission in the H$\alpha$ line. Indirect indications of the existence of a relationship between the total BD, $D=D^*+d$ (sometimes called anomaly of the BD), and the emission in the Balmer lines were also put forward by \citet{Feinstein1979} and \citet{Peton1981}. \citet{DivanZorecBriot1982} obtained a correlation between $d$ and the Balmer decrement. Since then scarce observational programs were carried out to study the appearance of the sBD. Apart from the models of CE by \citet{Poeckert1978a,Poeckert1978b}, where the sBD appears as a marginal phenomenon at extremely high CE densities, \citet{Moujtahid1999} and \citet{Cruzado2009} presented discussions on the behavior of the sBD and concluded that its appearance depends on the density distribution in the CE, its temperature, and inclination angle. \par

Knowing that the study of the appearance of the sBD could help us better understand the physical nature of the Be phenomenon, it is tempting to start a systematic study of the BD to uncover still unknown or perhaps unsuspected characteristics of the CE near the central star. Hence, the main goals of the present work are to analyze the frequency of appearance of the sBD as well as its aspect and intensity as functions of the stellar fundamental parameters in a sample of Be stars as large as possible. \par

In the Bright Star Catalogue \citep[BSC;][]{Hoffleit1982}, a non-negligible number of B and A stars are named Bn, An, but also Bnn and Ann stars. Their spectra are characterized by the hydrogen Balmer lines, lines of neutral helium, and lines of singly ionized oxygen, iron, and other gases that define the respective classical MK \citep{Morgan1973} B and A spectral type-luminosity classes. Most of these stars are in the B7-A2 range of spectral types. To the known MK spectral type designation, \citet{Adams1923} added the tag ``n'' to indicate that the spectroscopic lines (mostly metallic) are ``nebulous'' in contrast to sharper lines observed in other stars \citep{Ghosh1999}. In general, ``n'' stands for broad absorption lines, and ``nn'' for very broad absorption lines. It is presumed that the broad aspect of these lines is due to the stellar rapid rotation. In contrast to Be stars, Bn stars do not show any emission component in hydrogen and other lines. In a search for rapid $ubvy$ photometric variations in southern Be and Bn stars, \citet{Barrera1991} found that the amplitude of changes in Bn stars rarely exceed 0.02 mag. \par

Rapid rotation is a unifying property of the Be star group and certainly has a significant impact on the production of the Be phenomenon \citep[cf.][and references therein]{Zorec2016}. However, Bn stars also show very high rotation rates. If we assume that the broad and shallow spectral lines of Bn stars are due to rapid rotation, these stars would be observed equator-on. Rapidly rotating stars seen pole-on may also exist, but because of the lack of rotation-related features, we cannot identify these stars as rapid rotators. To our knowledge, there is no citation in the current literature to systematic research of gravity darkening-related (GD) signatures in the metallic lines of normal B-type stars with sharp lines, as is now being carried out for pole-on A-type stars (Royer et al. in preparation). \par

The frequency of Be stars has two maxima: one at B1-B2 spectral type and a secondary one at B5 spectral type \citep{ZorecBriot1997}, mostly as a consequence of the combination of the behavior of the source function of Balmer lines with the effective temperature and the shape of the initial mass function (IMF) \citep{Zorecetal2007}. Contrarily, Bn stars are mainly main-sequence (MS) stars of late B spectral type that seem to complete an apparent statistical lack of Be stars at these effective temperatures \citep{Zorec2000,Zorecetal2007}. Since only a few Bn stars have been observed with some emission in the Balmer lines \citep{Ghosh1999,Rivinius2006}, it is legitimate to ask whether Bn stars possess weakly developed circumstellar disks, or whether they have attained the required (unknown) physical properties to become full-fledged Be stars \citep{Zorecetal2005}. \citet{Gkouvelis2016} produced a simple model of circumstellar disk in Be stars to study the uncertainties affecting the aspect of the first BD due to the presence of the sBD and the intrinsic reddening of the Paschen continuum. As a by-product of this discussion, the authors show that the sBD can easily appear in late B-type stars with low-density CE. However, they are not able to raise sizeable emission signatures at least in the spectral lines, mainly H$\alpha$, to reveal the presence of a circumstellar disk. \par

Given these cited facts, we include Bn stars to the present study, not only as a function of fundamental parameters ($T_{\rm eff}$, $\log g$, $\log L/L_{\odot}$) which have been scarcely determined, but also to study these objects, on one hand in the frame of rapid rotation to search for links between Be and Bn stars, and on the other hand as incipient builders of CEs. We note the role of rotation in the incidence of the Be phenomenon. Based on a detailed analysis of rotational speeds for a large sample of Be stars, \citet{Zorec2016} conclude that most of these stars rotate at $V/V_{\rm c} \simeq 0.65$ and that the probability is low that these stars are critical rotators. This means that Bn stars probably do not need to become critical rotators to develop low-density CEs. However, there remains a question as to whether Be and Bn stars are differential rotators with critical or near-critical equatorial rotation, but we observe these stars as having an average global under critical linear velocity parameter $V$ because we get integrated radiation over the stellar hemisphere. \citep{Zorec2017a,Zorec2017b}. \par

In the present paper we present observational characteristics of Be and Bn stars mainly based on the behavior of their sBD and rotational velocities. A discussion on a possible stellar-class relationship between Be and Bn stars is then included. \par

Because the sBD is related to the presence of a CE, in a second paper of these series we will produce a model to test the required properties of the CE to give rise to this feature with its observed characteristics to uncover or emphasize additional links between Be and Bn stars. \par

\section{Observational data}\label{obsr}

Our sample consists of 66 Be and 61 Bn Galactic stars, listed in the BSC \citep{Hoffleit1982}. Low- and high-resolution spectroscopic observations for the sample stars were obtained at the Complejo Astronómico El Leoncito (CASLEO), San Juan, Argentina. In addition, we obtained H$\alpha$ spectra for some Be stars from the Be Star Spectra database \citep[BeSS;][]{Neiner2011} or \citet{Atlas}. We particularly searched for available spectra taken at dates as close as possible to that of our low-resolution observations. \par

In Tables \ref{tabla_Be} and \ref{tabla_Bn} we give, for Be and Bn stars, respectively, the HD number of each observed star, star coordinates, apparent visual magnitude, and dates of observation. For some targets we obtained observations in two different epochs. \par

All the low-resolution CASLEO observations were carried out using the Boller \& Chivens Cassegrain spectrograph attached to the Jorge Sahade telescope. Except for the 2004 spectra, the instrumental configuration consisted of a 600~$\ell$/mm grating, a slit width of 250 $\mu$m, and a TEK 1024 CCD detector. The covered spectral range was 3500-5000~\AA\, and the effective resolution 5.25~\AA\, every two pixels (R $\approx$ 660). For 2004 observations another CCD detector was used (PM 512), resulting in an effective resolution of 4.53~\AA\, every two pixels (R $\approx$ 900) and a little shorter spectral range (3500-4600~\AA). Accompanying high-resolution spectra were obtained using the Recherches et Études d’Optique et
de Science Connexes (REOSC) spectrograph, attached to the same telescope, in cross-dispersion mode with a 400 $\ell$/mm grating, a slit width of 250 $\mu$m, and a 1024$\times$1024 CCD detector. The observed spectral range is 4200--6700~\AA\, with \mbox{R $\approx$ 12600}. \par

The CASLEO observations were reduced using the {\sc IRAF}\footnote{IRAF is distributed by the National Optical Astronomy Observatory, which is operated by the Association of Universities for Research in Astronomy (AURA) under cooperative agreement with the National Science Foundation.} software package and applying a standard reduction procedure. Bias, flat-field, He-Ne-Ar comparison, and spectro\-photometric flux standard star spectra were secured to perform bias and flat-field corrections along with wavelength and flux calibrations for the low-resolution spectra. For the high-resolution spectra we used a Th-Ar Lamp as a comparison spectra. High-resolution spectra were not flux calibrated, and all the spectra were corrected for atmospheric extinction. \par

\begin{figure}
\centerline{\includegraphics[angle=-90,scale=0.7]{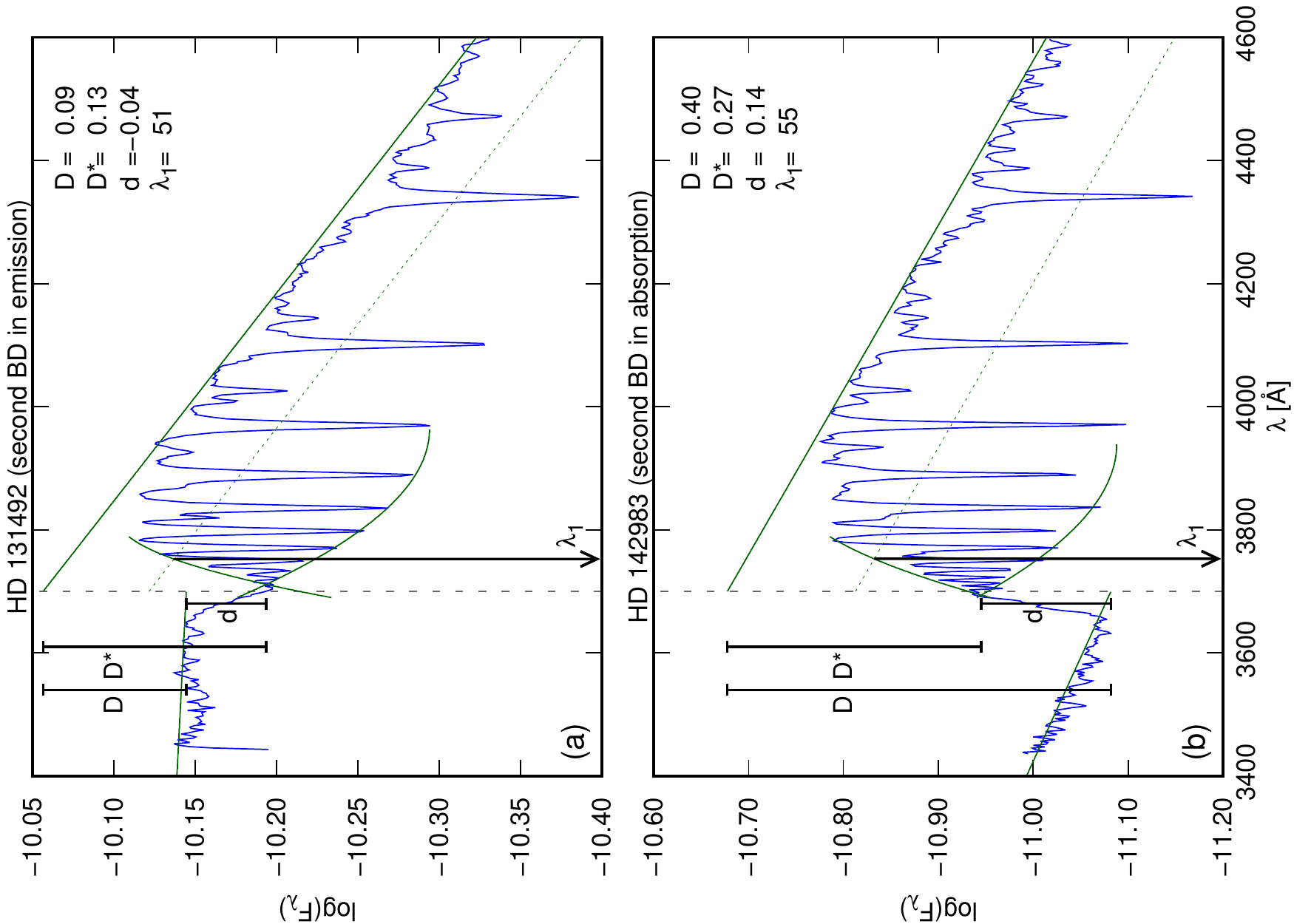}}
\caption{\label{secondBD} Low-resolution spectra for two different Be stars illustrating the cases for sBD in absorption and in emission. The different components of the BD are indicated: $D$ (total), $D^*$ (photospheric), and $d$ (circumstellar). The straight green lines at $\lambda>3700$~\AA\, and $\lambda<3700$~\AA\, represent the rectified Paschen and Balmer continuum flux. The vertical dashed line indicates the wavelength $\lambda~3700$~\AA\, at which the flux ratios are measured. The dotted green line represents $\log F_{+3700}-D^*/2$, and is used to determine the $\lambda_1$ parameter, whose abscissa is indicated with a vertical arrow (see Sect.~\ref{1stbd}.)}
\end{figure}

\subsection{BCD spectrophotometry}\label{sptrph}

\subsubsection{First component of the BD}\label{1stbd}

The BCD spectrophotometric classification system \citep[see basic definitions in][]{BarbierChalonge1941,ChalongeDivan1952,ZorecBriot1991} is based on measurements of the  BD in low-resolution spectra observed in the 3400--4600~\AA\,wavelength range. The photospheric component of the BD is characterized by the parameter $D^*$, which is given by the $\log$ of the ratio of the Paschen to Balmer continuum flux extrapolated at 3700~\AA\,, $D^*~=~\log(F_{+3700}/F_{-3700})$, and is a strong indicator of the stellar effective temperature, T$_{\rm eff}$. The flux $\log F_{+3700}$ is determined by extrapolating at $\lambda~3700$~\AA\,the rectified Paschen continuum flux $\log F_{\lambda}$. The value of $\log F_{-3700}$ is determined by the conjunction of the upper and lower wraparound curves of the last absorption lines of the Balmer series, as shown in Fig.~\ref{secondBD}. The parameter $\lambda_1$ is the mean spectral position of the Balmer jump, relative to $\lambda\ 3700$\,\AA, and provides a good indication of the stellar surface gravity, $\log g$. It is determined by the intersection of the line parallel to the Paschen continuum that passes through the point \mbox{(3700, $\log {F_{+3700}-D^*/2}$)} (dotted line in Fig.~\ref{secondBD}) and the upper wraparound curve of the last members of the Balmer line series. \par

The measurement of the BCD parameters was performed with an interactive code developed by one of us (Y.A.) and successfully implemented to study B stars in open clusters \citep{Aidelman2012,Aidelman2015,Aidelman2018}. Typical uncertainties of these quantities are $\sigma(\rm{D}^*)\!\la\!0.015$ dex and $\sigma (\lambda_1)\!\sim\!3$\AA. The BCD $(\lambda_1, D^*)$ parameters allow us a straightforward determination of the spectral type and the luminosity class of stars, as well as of the stellar parameters \citep{Zorecetal2009}. In general, the spectral types based on the BCD stellar parameters of emissionless stars are in agreement with the MK classification within one to two subtypes. \par

Tables~\ref{tabla1_Be} and \ref{tabla1_Bn} list the ($\lambda_1, D^*$) parameters in Cols. 2 and 3, and the corresponding spectral types and luminosity classes determined in this work (Col. 4) of the program Be and Bn stars, respectively. Six objects of our sample: \object{HD 31209}, \object{HD 42327}, \object{HD 43445}, \object{HD 165910}, \object{HD 171623,} and \object{HD 225132}, considered up to now as Bn stars, turned out to have an emission component in their H$\alpha$ line, so they are included in the Be star sample. \par

\begin{figure}
\centering
\includegraphics[width=\hsize]{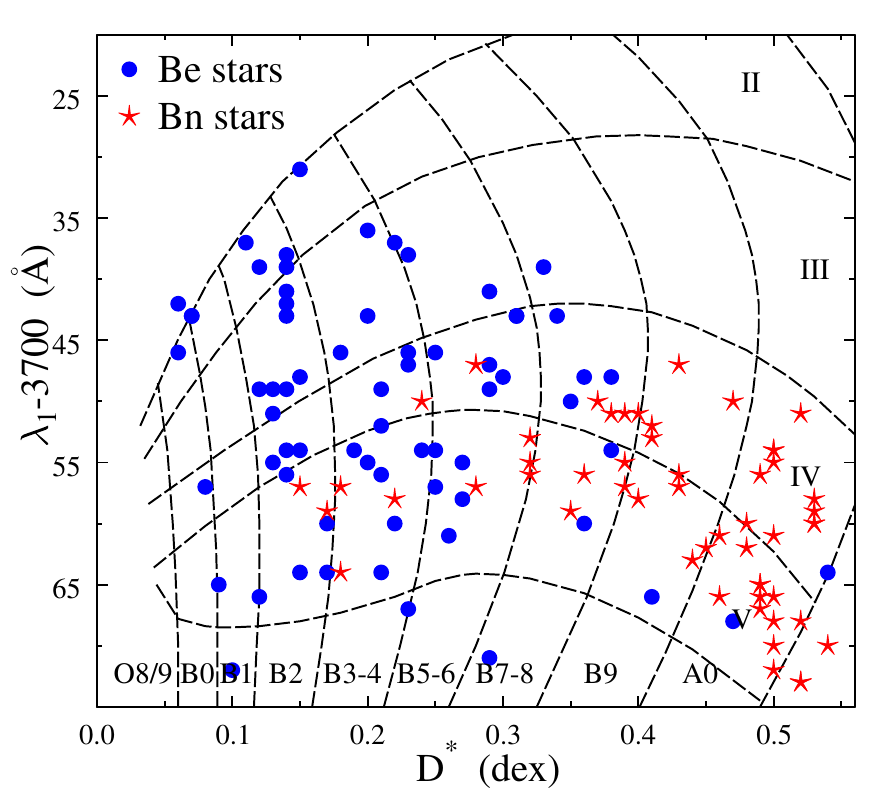}
\caption{Calibration of the BCD $(\lambda_1, D^*)$ plane in terms of the MK spectral type. The vertical dashed lines show the limits of the MK spectral type groups indicated on the axis of abscissas. The horizontal dashed lines represent the limits of the MK luminosity classes, which are indicated in the last vertical curvilinear quadrilaterals. Our sample of Be stars has luminosity classes in the V-II range, while the Bn stars show luminosity classes V or IV.}
\label{bcdl1a}
\end{figure}

Figure~\ref{bcdl1a} shows the BCD $(\lambda_1, D^*)$ version of the Hertzsprung-Russell (HR) diagram of the studied Be and Bn stars. This diagram shows that the BCD parameters ($\lambda_1, D^*$) have high precision in both spectral type and luminosity class. In some particular cases, when emission is very strong, mostly among the hottest stars (from O8- to B1-type), the BCD spectral types can be somewhat earlier than expected (around half spectral type). This happens because the emission in the sBD can slightly overlap the photospheric emission, when the continuum flux excess and the emission in the spectral lines are very strong. However, the separation between the photospheric and circumstellar Balmer discontinuities is clear for all the remaining cooler spectral types. On one hand, because of this separation and because the parameter $D^*$ is determined by a ratio of fluxes at the same wavelength ($\lambda=3700$~\AA), this quantity is independent of the circumstellar perturbing emission and absorption, and its dependence with the interstellar extinction is only of second order, through the slope of the Paschen energy distribution, $\log F_{\lambda}$, which is on the order of $\delta D^* \lesssim 0.01E(B-V)$ dex. On the other hand, the $\lambda_1$ parameter is independent of both the interstellar extinction and the circumstellar absorption and emission. Our sample of Be stars has luminosity classes in the V-II range, while the Bn stars show luminosity classes V or IV. \par

\begin{figure}
\centering
\includegraphics[width=0.48\textwidth]{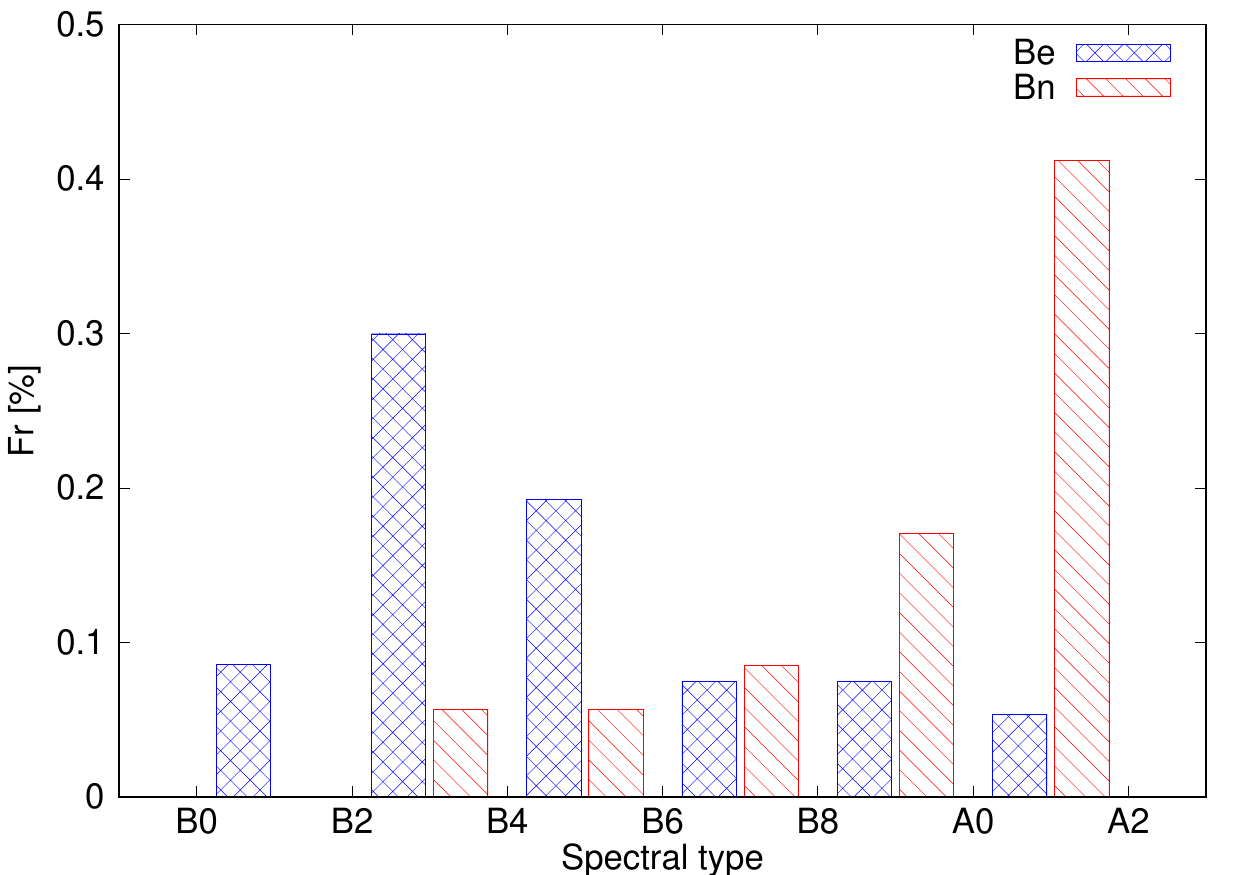}
\caption{Histogram of BCD spectral types of the studied Be and Bn stars. The maximum frequency of Be stars is located around the B2 spectral type. The number of Bn stars strongly increases toward late B and early A-type spectral types. Both Be and Bn stars spread over the entire range of B-type stars.}
\label{histo}
\end{figure}

\begin{figure} 
\centering
\includegraphics[width=0.35\textwidth]{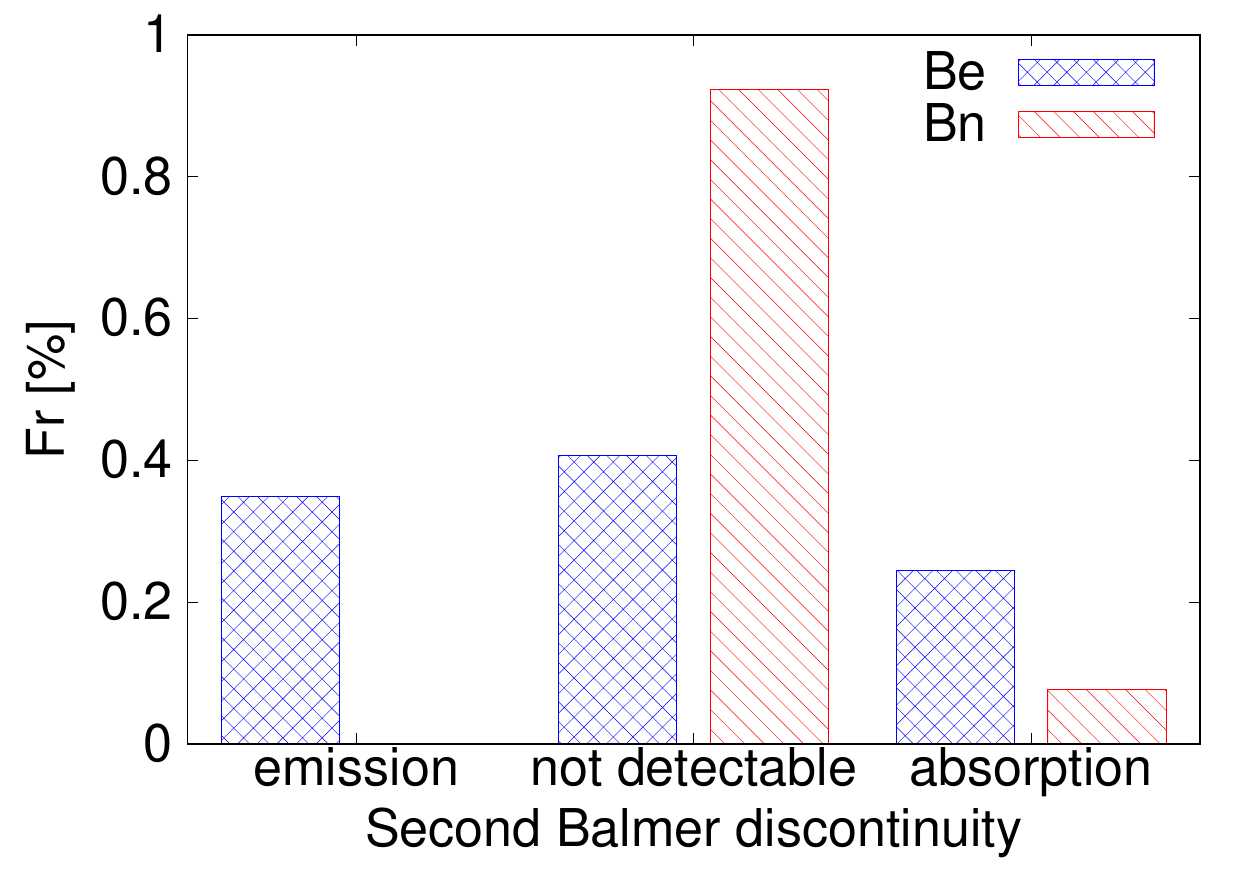}
\caption{Diagram showing the frequency of appearance of the sBD in our Be and Bn stellar sample. More than half of the Be stars in our sample present the sBD either in emission or in absorption. The fraction of Bn stars that present the sBD in absorption is small.}
\label{secondBDstd} 
\end{figure}

Figure~\ref{histo} shows the occurrence histogram of the frequency of spectral types in the Be and Bn stellar sample studied using the BCD system. In this figure the maximum frequency of Be stars is located around the B2 spectral type, in agreement with that reported by other authors for other samples of Be stars \citep[e.g.,][]{Jaschek1983,ZorecBriot1991}. As already noted in the introduction, the number of Bn stars strongly increases toward late B and early A-type spectral types. Nevertheless, it is interesting to note that both Be and Bn stars spread over the entire range of B-type stars. \par

\subsubsection{Second Balmer discontinuity}\label{2ndbd}

\begin{figure*}[h!]
\centering
\includegraphics[angle=-90,scale=0.67]{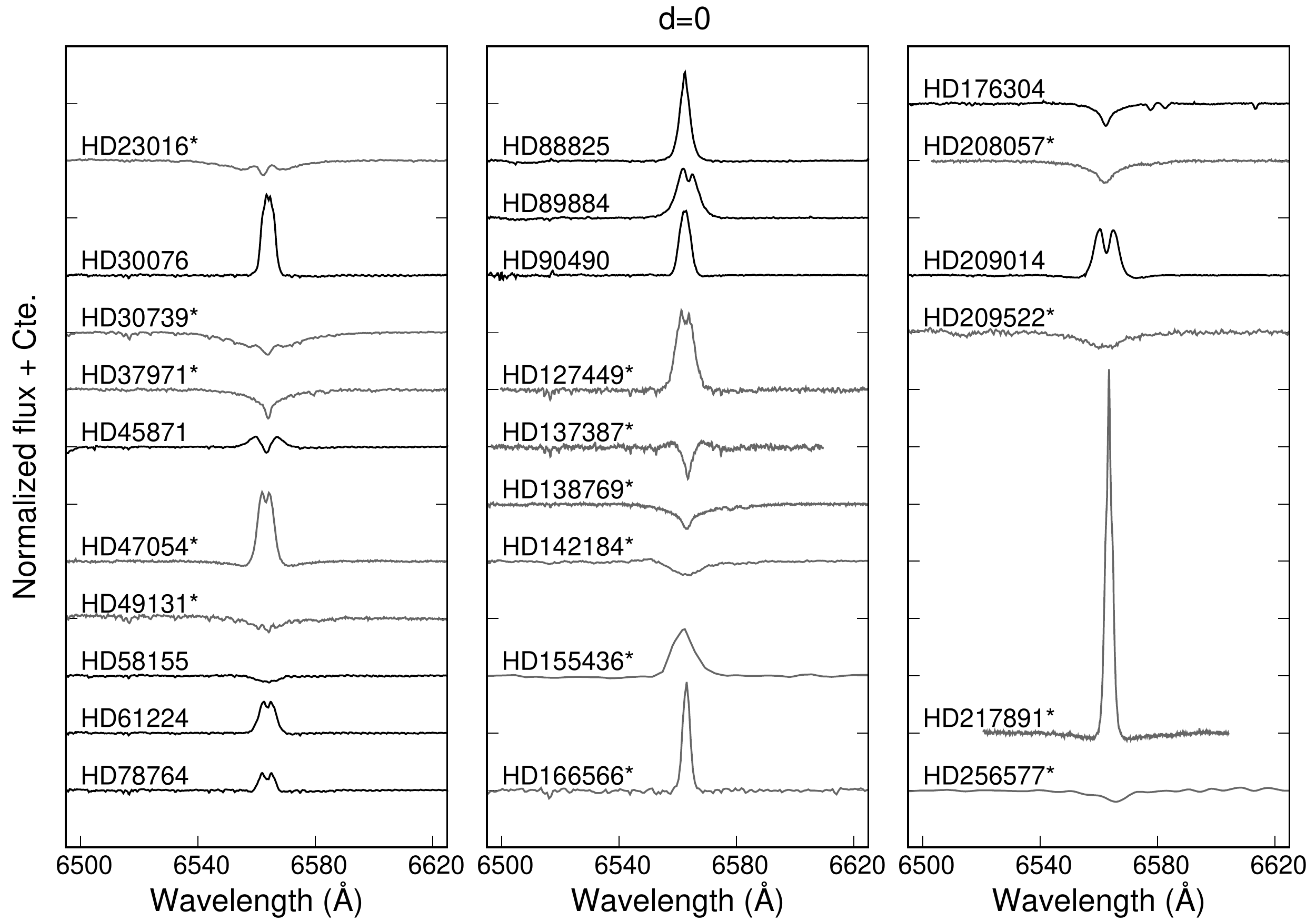}
\caption{H$\alpha$ line profiles belonging to the program Be stars without a sBD, observed at dates close to that of our low-resolution observations. Spectra with the * symbol were downloaded from the BeSS database \citep[][]{Neiner2011} or \citet{Atlas}.}
\label{halphalines_0} 
\end{figure*}

One of the objectives pursued in this work is to characterize the sBD component in a rather large sample of Be and Bn stars. In general, Be stars can have a BD with two components. One of these components ($D^*$, mentioned in Sect.~\ref{1stbd}) characterizes the photosphere of the rotationally deformed stellar hemisphere projected toward the observer, and the second ($d$) is produced by the circumstellar gaseous environment of these objects. We refer then to a total BD written as $D = D^* + d$ dex. According to the definition put forward by \citet{Divan1979}, when the sBD is in emission we have $d<0$, and $d>0$ when it is in absorption. In Fig.~\ref{secondBD} we show examples of spectra with the sBD in emission (top) and in absorption (bottom), and where the determination of $D$, $D^*$, and $d$ is indicated. \par

Once those Bn stars for which we found an emission component in the H$\alpha$ line were moved to the Be star group, the Be and Bn star samples were left with 73 and 55 objects, respectively. Among the 73 Be sample stars, 38 have a measurable sBD, 23 of which are in emission and the other 15 in absorption. In 13 spectra of the whole sample, we see an incipient sBD, but it is too small to be measured reliably. Seven of these spectra present $d \lesssim 0$ and the other six $d \gtrsim 0$. In all remaining Be stars only the first BD is seen. Regarding the Bn stars, 3 out of 55 stars show a measurable sBD in absorption, and other two spectra present $d \gtrsim 0$, but they are too tiny to be measured with certainty. In the remaining spectra of Bn stars there is no detectable sBD. A graphical summary of the fraction of stars versus the aspect of the sBD for Be and Bn stars is presented in Fig.~\ref{secondBDstd}. \par

The measured values of the sBD, $d$, are given in Col. 5 of Tables~\ref{tabla1_Be} and \ref{tabla1_Bn} for Be and Bn stars, respectively, where $d$ values range from -0.07 dex to 0.14 dex for Be stars and from 0.00 dex to 0.03 dex for Bn stars. For those objects with two measurements of $d$, we present both values. The text ``em'' or ``abs'' in the $d$ column indicates those cases in which the sBD is too small to be measured reliably. \par

\begin{figure*}
\centering
\includegraphics[angle=-90,scale=0.67]{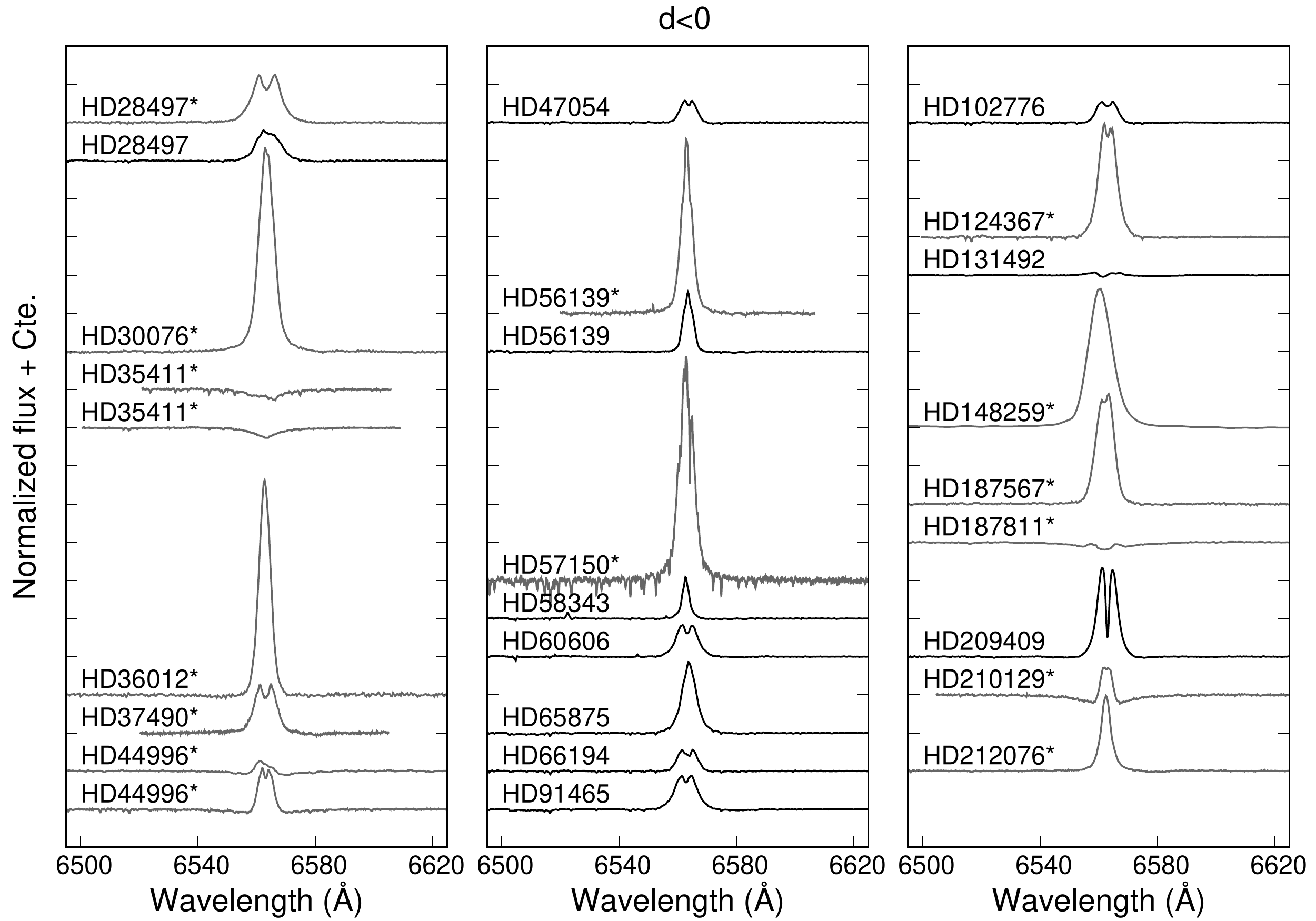}
\includegraphics[angle=-90,scale=0.67]{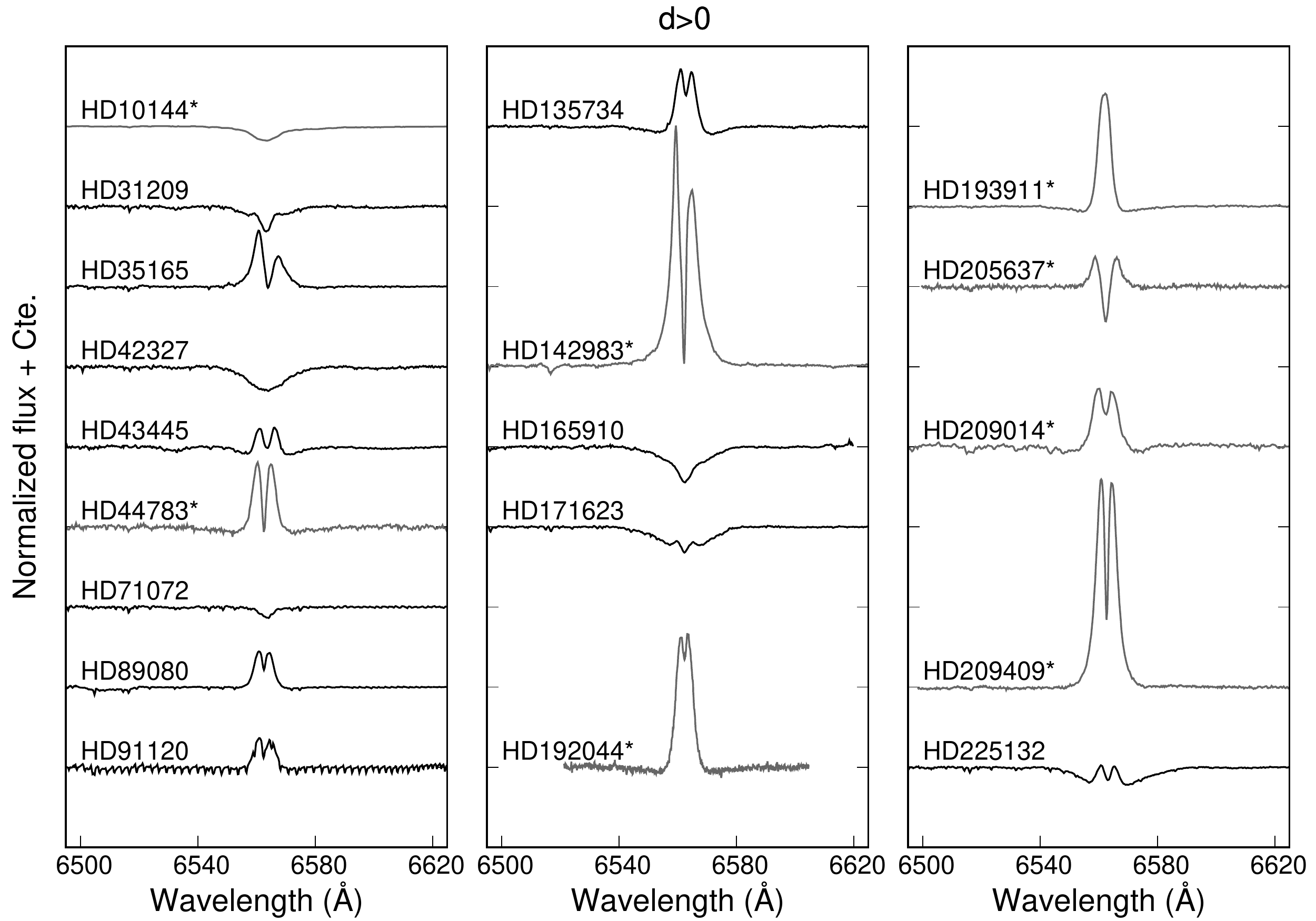}
\caption{H$\alpha$ line profiles belonging to the program Be stars with sBD in emission ($d < 0$, top panel) and in absorption ($d > 0$, bottom panel), observed at dates close to those of our low-resolution observations. Spectra with the * symbol were downloaded from the BeSS database \citep[][]{Neiner2011} or \citet{Atlas}.}
\label{halphalines_emab}
\end{figure*}

\subsection{H$\alpha$ line observations for program Be and Bn stars}\label{newBe}

\subsubsection{Be stars}\label{habe} 

In this section we present our H$\alpha$ observations for the program Be stars. Figures~\ref{halphalines_0} and \ref{halphalines_emab} show the H$\alpha$ line profiles for the stars without and with a sBD, respectively. In Fig.~\ref{halphalines_emab}, the top panel shows the spectra for objects with $d<0$ while the lower panel corresponds to those with $d>0$, including the stars for which the sBD is too small to be measured with certainty. In both figures, the symbol * stands for those spectra taken from the already mentioned databases. We note that HD 209014 has been seen once with $d>0$ and another time without a sBD. \par

In our search for Be stars that show a clear sBD, we have preferentially chosen those with strong H$\alpha$ emission, based on the correlations given in \citet{DivanZorecBriot1982}. Unfortunately, this may contribute with some biased effects on the aspect angle statistics carried in later sections of this work. Indeed, we selected stars with rather low aspect angles, which likely display stronger emissions due to the CE. \par

It is worth mentioning that for \object{HD\,35411}, the upper H$\alpha$ line profile in Fig.~\ref{halphalines_emab}, that presents a small emission above the photospheric absorption, have been taken around three years before our first low-resolution spectrum with the sBD in emission. Also, the other H$\alpha$ line profile is for the same month of our second low-resolution spectrum, but does not show evidence for an emission component, expected due to the sBD in emission. It is possible that the development of a CE was recent. On the other hand, \object{HD\,187811} shows a H$\alpha$ line profile with a weak double-peaked shell-like emission overimposed to the photospheric absorption. This spectra was taken almost four months before our low-resolution spectrum with a sBD in absorption. Considering the spectral variability of this object \citep{Mennickent2009,Vieira2017,Sabogal2017}, it is possible that the small emission in the H$\alpha$ line occurred owing to the development of a new CE that was present when the sBD was observed. \par

In Fig.~\ref{halphalines_emab} ($d>0$, bottom panel) the H$\alpha$ profiles of the six newly discovered Be stars, which were previously classified Bn stars, are included. For \object{HD\,31209} and \object{HD\,171623}, the H$\alpha$ line profiles has a shell-like double-peaked weak emission cutting the bottom of the underlying photospheric absorption. In \object{HD\,43445} and \object{HD\,225132}, the H$\alpha$ profiles show shell-like intense double-peaked emission superimposed on the photospheric absorption. In \object{HD\,42327} we see a particularly rounded bottom of the line absorption possibly due to some emission that partially fills up the underlying photospheric absorption component. In \object{HD\,165910, however,} there is an extra absorption in the bottom of the photospheric component, which is typical for B-shell stars \citep{Hubert1979}. \par 

\subsubsection{Bn stars}\label{habn} 

\begin{figure*}[h!] 
\centering
\includegraphics[width=0.23\textwidth]{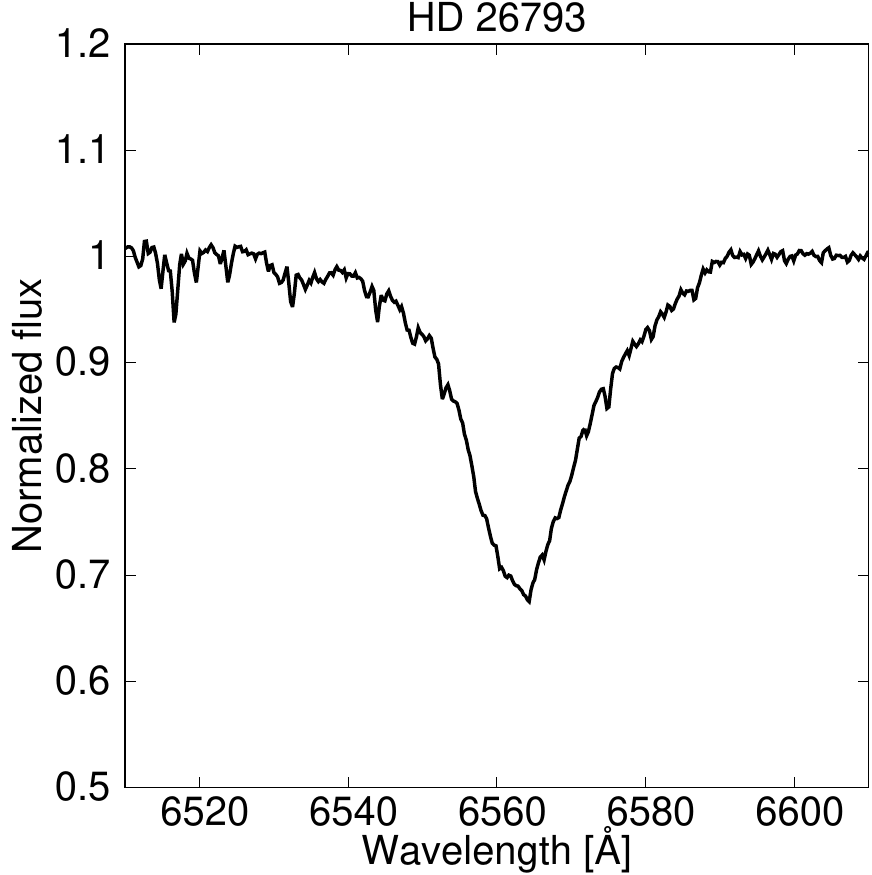}
\includegraphics[width=0.23\textwidth]{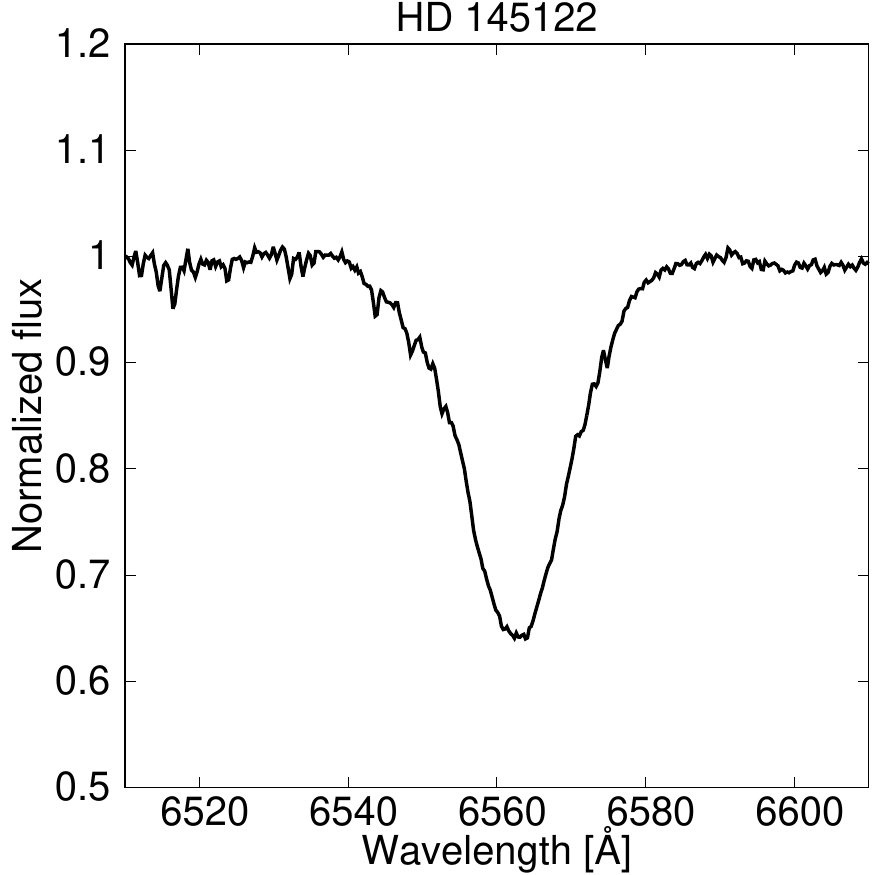}
\includegraphics[width=0.23\textwidth]{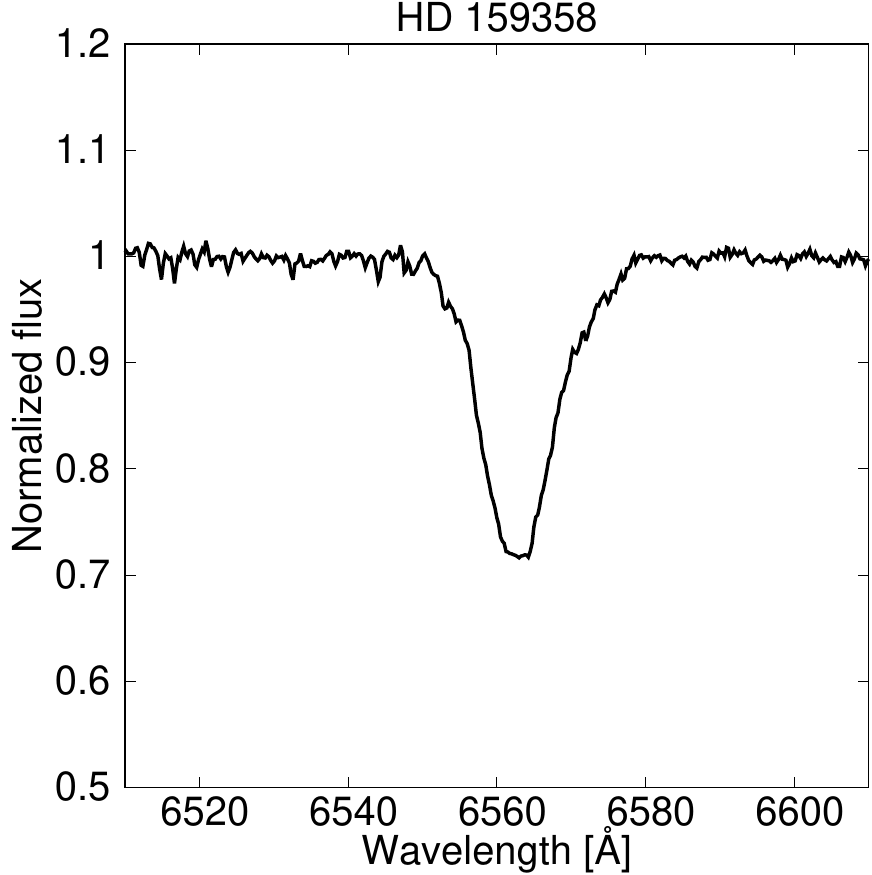}
\includegraphics[width=0.23\textwidth]{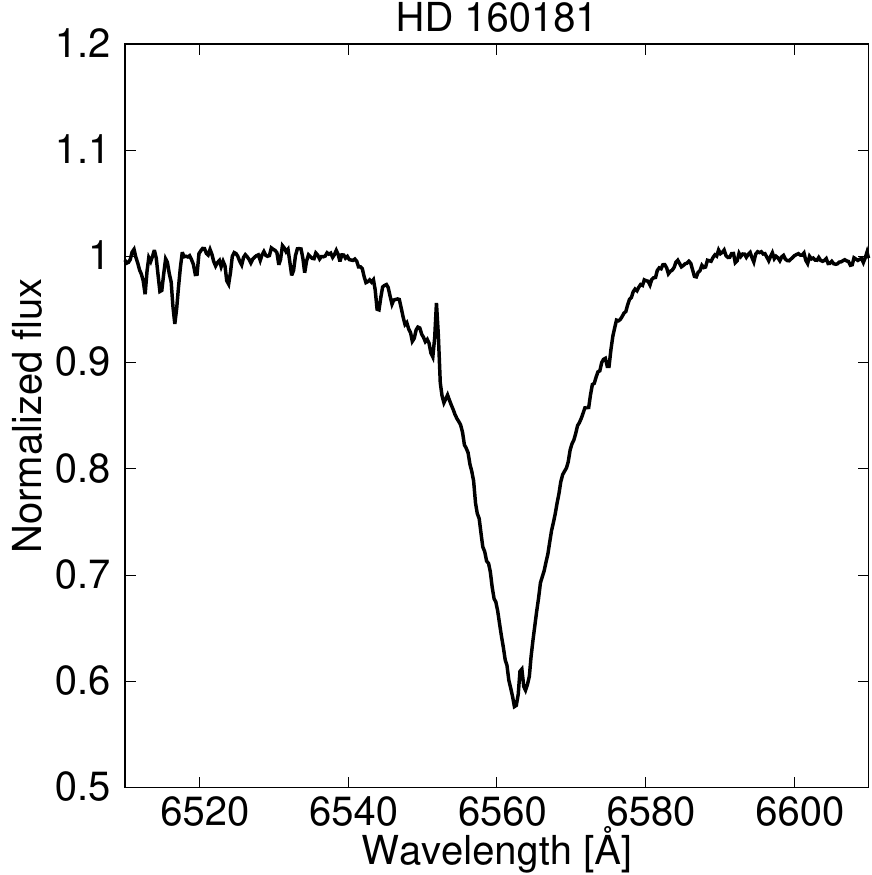}
\includegraphics[width=0.23\textwidth]{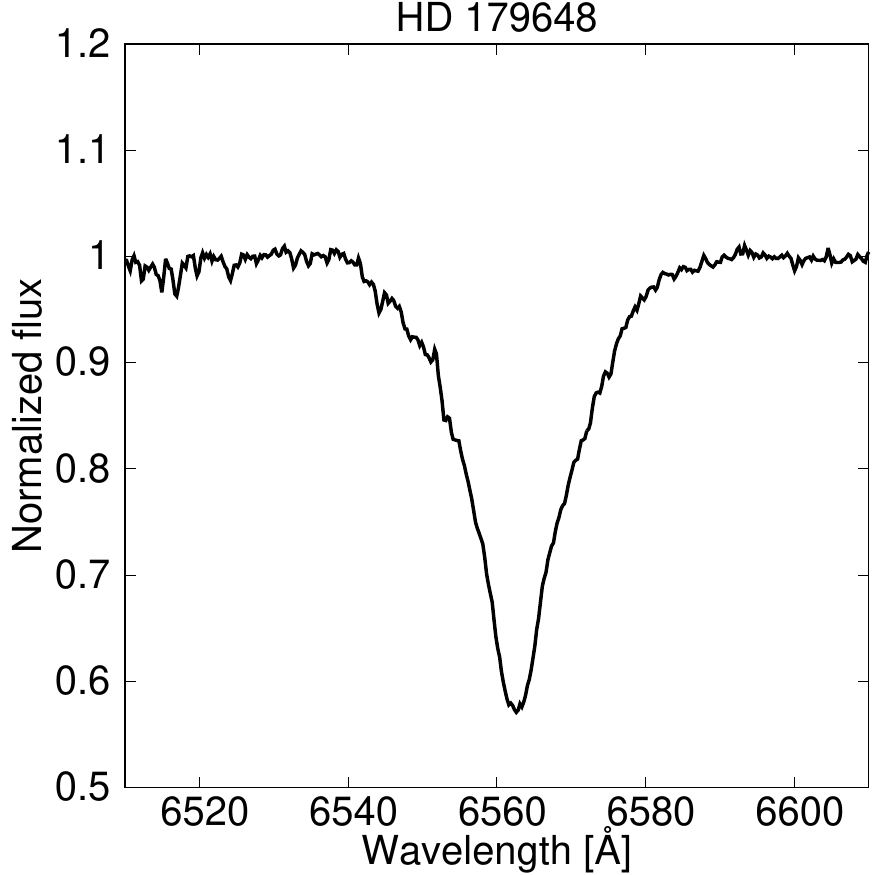}
\includegraphics[width=0.23\textwidth]{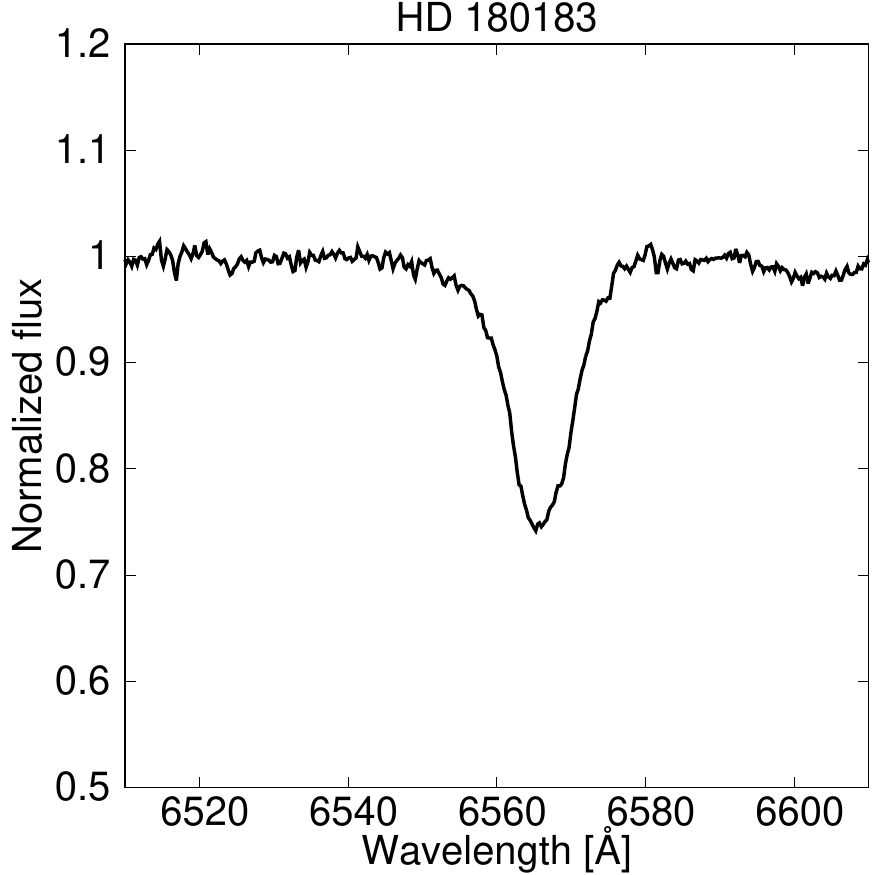}
\includegraphics[width=0.23\textwidth]{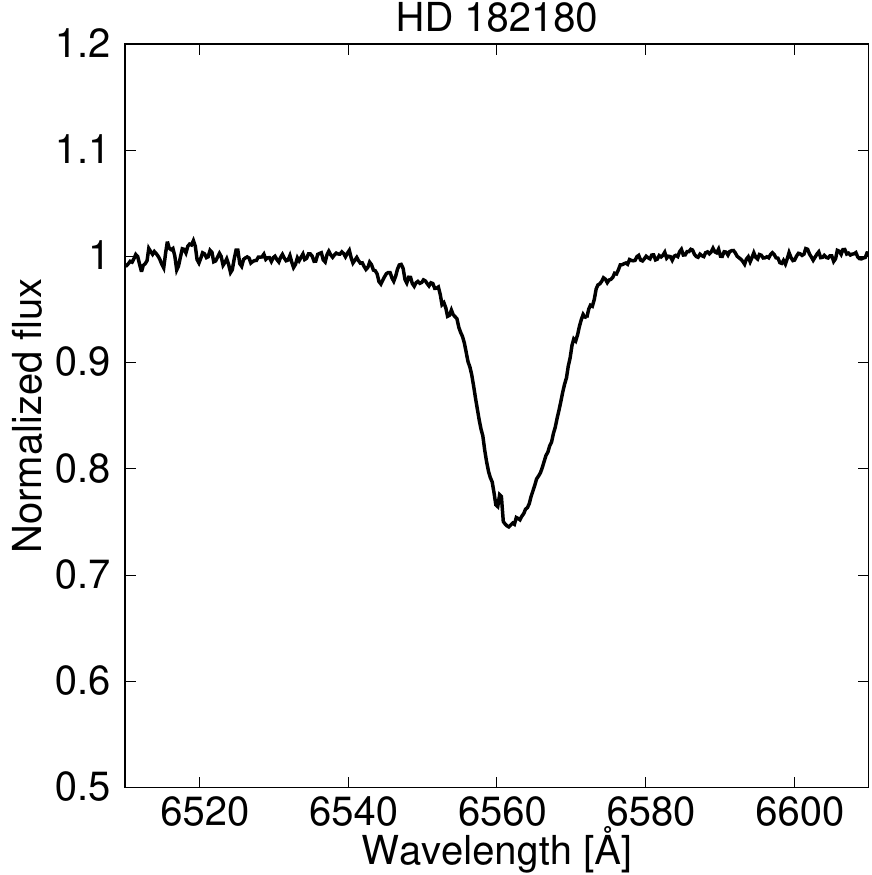}
\includegraphics[width=0.23\textwidth]{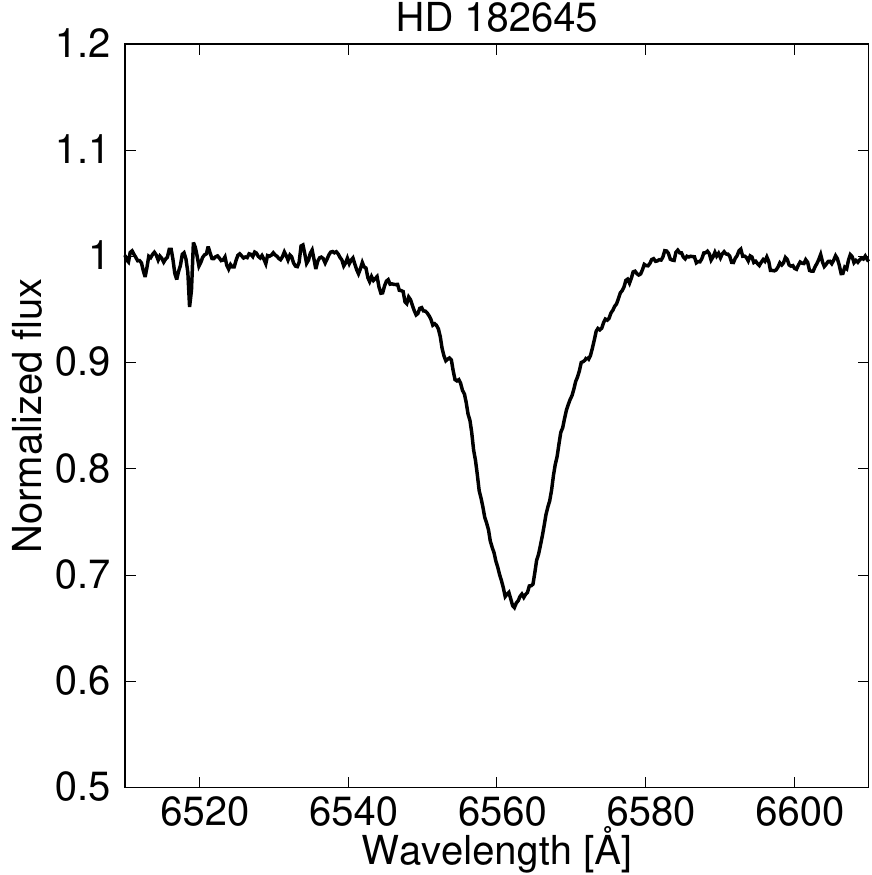}
\includegraphics[width=0.23\textwidth]{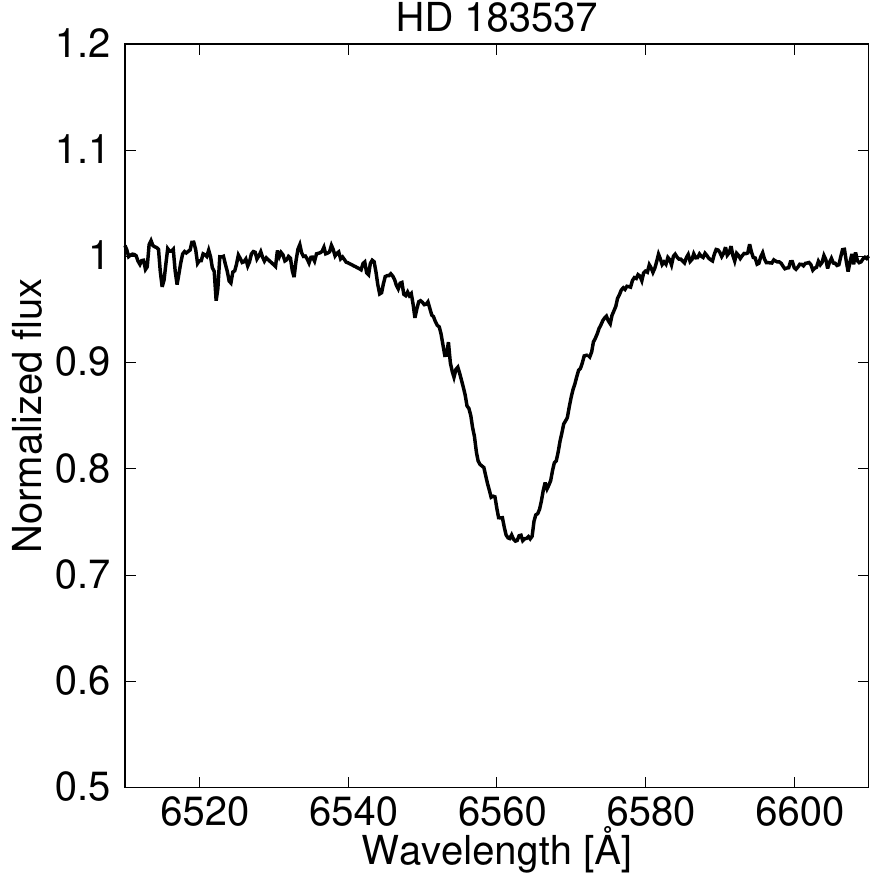}
\includegraphics[width=0.23\textwidth]{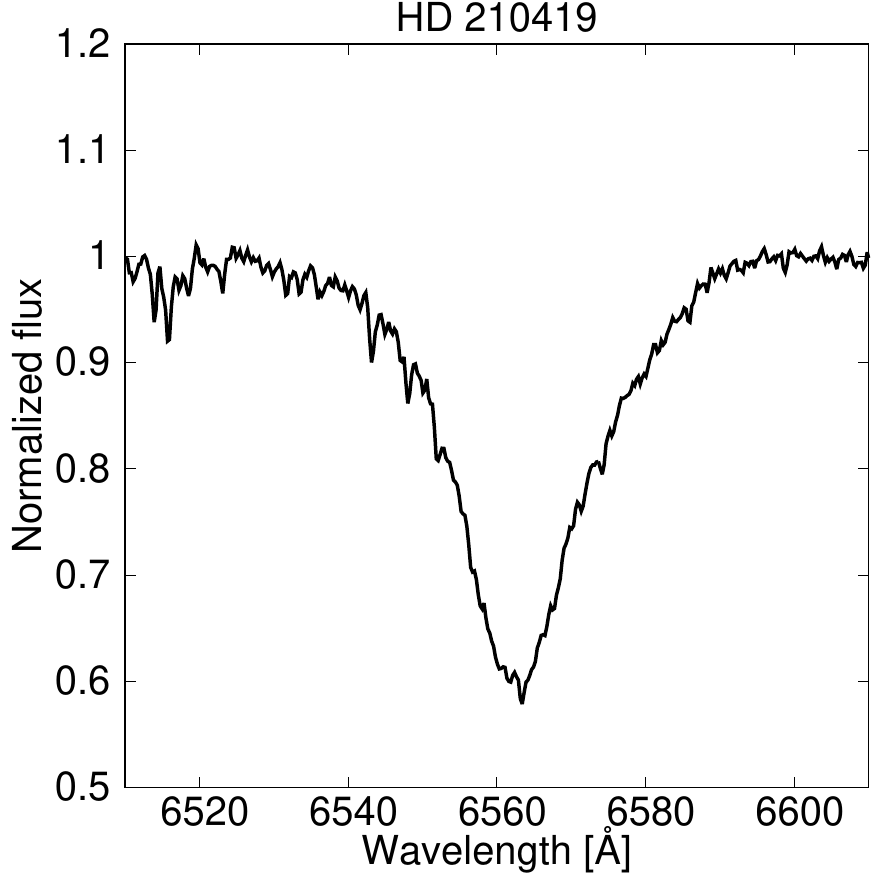}
\caption{H$\alpha$ line for Bn stars that present a sBD, including the stars for which the sBD is too small to be measured with certainty. These profiles have a photospheric-like absorption aspect.}
\label{Ha_Bn}
\end{figure*}

We obtained high-resolution H$\alpha$ spectra for those program Bn stars that present (or might present) a sBD. Since the sBD is most probably originated by a more or less extended envelope around the star, we expected to detect emission in the H$\alpha$ line as well. \par

As mentioned in Sec.~\ref{1stbd}, six stars previously classified as Bn stars have been included in the Be star group because they show emission components or shell-like absorptions in their H$\alpha$ line profiles. These stars are considered in this paper as genuine Be stars and were included in the Be star group. \par

In Fig.~\ref{Ha_Bn} we show the H$\alpha$ line profiles for those Bn stars with an indication of a possible sBD in their low-resolution spectra. All the line profiles have a photospheric-like aspect. However, to determine whether they are affected by some emission or extra absorption produced by the CE, it is still necessary to fit theoretical profiles to simultaneous high-resolution observation in the blue (H$\gamma$, H$\delta$, H$\epsilon$ lines) and red (H$\alpha$) spectral ranges. Differences between observed and modeled H$\alpha$ line profiles could finally determine whether there is some CE emission or extra absorption. \par

The rapid rotation that characterizes Bn stars, the occasional appearance of some emission in the Balmer lines (at least in H$\alpha$), and the presence of signatures for some sBD are motivate us to think that Bn and Be probably belong to the same class of objects. Some attempts to show a possible correlation between fast-rotating late B-type stars and Be stars based on the presence of a sBD were discussed by \citet{Aidelman2018} in open clusters. \par

\subsection{$V\!\sin i$ parameters}\label{vsini}

With the purpose of searching for physical reasons that could justify class relations between Be and Bn stars, we compiled $V\!\sin i$ parameters of the program stars published in the literature or determined in this work. The $V\!\sin i$ values were taken from \citet{Chauville2001}, \citet{Zorecetal2005}, \citet{Fremat2005} and \citet{Zorec2012}. We also adopted values from \cite{Dworetsky1974}, \citet{Slettebak1982}, \citet{Uesugi1982}, \citet{Wolff1982}, \citet{Brown1997}, \citet{Yudin2001}, \citet{Abt2002}, \citet{Strom2005}, \citet{Levenhagen2006} and \citet{Diaz2011}. For three of the program stars we did not find any $V\!\sin i$ determinations nor the required spectra to be able to measure them. The finally adopted values of $V\!\sin i$ and their uncertainties are listed in columns 13 and 14 of Tables~\ref{tabla1_Be} and \ref{tabla1_Bn}. Column 15 of these tables indicates the respective references. \par

\section{Second Balmer discontinuity and the inclination angle of the rotation axis}\label{sbdi}

\begin{figure}
\centerline{\includegraphics{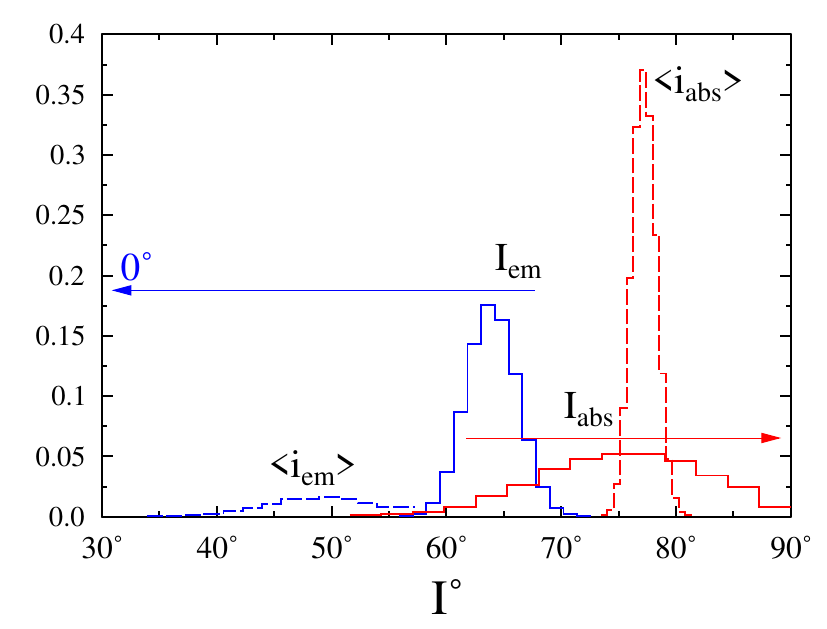}}
\caption{\label{fig_incl} Full lines: Histograms of Monte Carlo simulations for limit inclinations $I_{\rm em,abs}$ defined in Eq.~\eqref{eq_1}. Dashed lines: Histograms of average angles $\langle i_{\rm em,abs}\rangle$ defined in Eq.~\eqref{eq_2}. Blue indicates stars with sBD in emission, red indicates sBD in absorption. The arrows indicate the domains over which the integrations that define $\langle i_{\rm em,abs}\rangle$ were performed, respectively.}
\end{figure}

In several occasions, evidence was put forward to claim that flux excesses representing either emission or absorption in the visible continuum spectrum of Be stars should be produced in circumstellar disk layers laying close to the central star \citep{Moujtahid1998,Moujtahid1999,Moujtahid2000}. This suggestion seems to be supported by recent calculations of the visual energy distribution in Be stars based on models of viscous CE \citep{Carciofi2008,Haubois2012,Klement2017}. \par

Knowing that most Be stars show a sBD either in emission or absorption, we are tempted to ask whether there is a relationship between the inclination angle and the geometry of the CE. Nevertheless, it is worth noting that a few cases were reported in the literature in which the genuine Be aspect of the H$\alpha$ line remained unchanged over a long period of time and then changed to a genuine Be-shell aspect or the line emission or extra absorption component simply disappeared for a while. The sBD might then also change from emission to absorption or simply disappear \citep{HirKog1977,Moujtahid1999}. In these particular cases, finding this kind of relationship might be somewhat challenging. \par

In this section, we carried out statistical tests to determine the observational circumstances under which the sBD appears by trying to infer the ranges of inclination angles at which it may appear in emission or in absorption. Knowing that the $V\!\sin i$ parameters are currently underestimated owing to the GD effect \citep{Stoeckley1968,Townsend2004,Fremat2005,Zorec2016}, to obtain a rough first insight on the inclination angles we  corrected the projected rotational velocities, assuming that all the studied stars rotate on average at the same angular velocity ratio $\Omega/\Omega_{\rm c}=0.95$; this ratio closely corresponds to the average ratio $\eta=0.6$ of equatorial centrifugal to gravitational forces characterizing a large sample of Be stars near the Sun \citep{Zorec2016}. \par

We call $\langle V\!\sin i\rangle_{\rm em}$ the average $V\!\sin i$ parameter that corresponds to those program Be stars with the sBD in emission, and $\langle V\!\sin i\rangle_{\rm abs}$ the average characterizing the Be stars with the sBD in absorption. Admitting that the circumstellar gaseous structures around Be stars are globally flattened, we can think of a sBD in emission from objects that are preferentially seen at inclination angles from $i=0\degr$ to some $i=I_{\rm em}$, while the sBD in absorption is detected in stars having aspect angles from some $i=I_{\rm abs}$ to $i=90\degr$. The distribution function of the actual linear velocity $V$ and that for the inclination angle $i$ are independent. Since $\sin i\,di$ is the probability of finding an object inclined between the angles $i$ and $i+di$, the $\langle V\!\sin i\rangle_{\rm em,abs}$ averages are formally given by

\begin{equation}
\begin{array}{rcl}
\displaystyle \langle V\!\sin i\rangle_{\rm em} & = & \displaystyle \langle V\rangle\frac{\int_{0}^{I_{\rm em}}\sin^2i\,di}{\int_{0}^{I_{\rm em}}\sin i\,di} \\
& = & \displaystyle \frac{\langle V\rangle}{2}\left(\frac{I_{\rm em}-\frac{\sin 2I_{\rm em}}{2}}{1-\cos I_{\rm em}}\right) \\

\displaystyle \langle V\!\sin i\rangle_{\rm abs} & = & \displaystyle \langle 
V\rangle \frac{\int_{I_{\rm abs}}^{\pi/2}\sin^2i\,di}{\int_{I_{\rm abs}}^{\pi/2}\sin i\,di} \\
& = & \displaystyle \frac{\langle V\rangle}{2}\left(\frac{\frac{\pi}{2}-I_{\rm abs}+\frac{\sin 2I_{\rm abs}}{2}}{\cos I_{\rm abs}}\right),
\end{array}
\label{eq_1}
\end{equation}

\noindent where $\langle V\rangle$ is the average of all true rotational velocities of the entire Be star sample. In a similar way, the average angles $\langle i_{\rm em}\rangle$ and $\langle i_{\rm abs}\rangle$ under which are seen the sBD in emission or absorption are functions of $I_{\rm em}$ and $I_{\rm abs}$, respectively, given by

\begin{equation}
\begin{array}{rcl}
\displaystyle \langle i_{\rm em}\rangle & = & \displaystyle \frac{\int_{0}^{I_{\rm em}}i\,\sin i\,di}{\int_{0}^{I_{\rm em}}\sin i\,di} = \displaystyle \frac{\sin I_{\rm em}-I_{\rm em}\cos I_{\rm em}}{1-\cos I_{\rm em}} \\

\displaystyle \langle i_{\rm abs}\rangle & = & \displaystyle \frac{\int_{I_{\rm abs}}^{\pi/2}i\,\sin i\,di}{\int_{I_{\rm abs}}^{\pi/2}\sin i\,di} = \displaystyle \frac{1+I_{\rm abs}\sin I_{\rm abs}}{\cos I_{\rm abs}}.
\end{array}
\label{eq_2} 
\end{equation} 

In Tables~\ref{tabla1_Be} and \ref{tabla1_Bn}, the adopted apparent $V\!\sin i$ values with their uncertainties as reported in the literature are given in Cols. 13 and 14 \citep[for definition of apparent fundamental parameters, see][]{Fremat2005,Zorec2016}. Assuming that these uncertainties correspond to the standard deviation of the individual estimates, we carried out $10^4$ Monte Carlo bootstrapped trials to calculate the averages $\langle V\!\sin i\rangle_{\rm em,abs}$ defined in Eq.~\eqref{eq_1} and estimated the respective inclination angles $I_{\rm em}$ and $I_{\rm abs}$ and their corresponding averages given in Eq.~\eqref{eq_2}. The normalized distributions of $I_{\rm em,abs}$ and $\langle i_{\rm em,abs}\rangle$ thus obtained are shown in Fig.~\ref{fig_incl}. In this figure, the histograms in full lines show that the values of $I_{\rm em}$ and $I_{\rm abs}$  spread over large intervals of inclination angles owing to the uncertainties of the individual $V\!\sin i$ parameters. The histograms drawn with dashed lines correspond to the distributions of $\langle i_{\rm em}\rangle$  and $\langle i_{\rm abs}\rangle$  that indicate at what angles we can expect to observe the sBD in emission and in absorption, respectively. The arrows indicate the domains over which the integrations that define $\langle i_{\rm em,abs}\rangle$ were performed, respectively. \par

The results shown in Fig.~\ref{fig_incl} indicate that according to the observed averages $\langle V\!\sin i\rangle_{\rm em}$ there must be stars with a sBD in emission observed at inclination angles $I_{\rm em}\lesssim 70\degr$, while the absorption in the sBD should be displayed in stars that have inclination angles $I_{\rm abs} \gtrsim 60\degr$. Since the extra emission or absorption in the continuum spectrum produced by the CE detected as a sBD cannot come from circumstellar layers situated too far from the central star, our statistical inferences suggest that the CEs must have high enough vertical optical depths in these regions; this is very similar to suggestions made sometime ago by \citet{Arias2006} and \citet{Zorec2007_470} to account for the presence of \ion{Fe}{ii} emission lines originating in CE layers that are near the central star. The rather significant vertical height of CE layers near the star is also supported by the fact that rarely the sBD appears at wavelengths that are larger than the theoretical limit of Balmer lines series, $\lambda3648$\,\AA\,\citep{Divan1979,ZorecBriot1991,Moujtahid1999,Gkouvelis2016}, where the electron density of the CE layers close to the central star cannot be larger than some $N_{\rm e}\sim10^{13}$ cm$^{-3}$. Otherwise the sBD would heavily encroach upon the first or photospheric BD, phenomenon that was exceptionally observed during some emission phases. Thus, to attain optical depths $\tau \lesssim 1$ in the visible continuum spectrum in CE layers close to the central star that produce a sBD, the CE should have a high enough vertical height. \par

\section{Apparent and parent nonrotating counterpart parameters}\label{apnrc}

\subsection{Apparent parameters}\label{apppar}

In our vocabulary, the observed quantities are considered as ``apparent'' parameters. The observed BCD ($\lambda_1, D^*$) quantities are then apparent, but also all those which are derived or read in their calibrations, like the effective temperature $T_{\rm eff}(\lambda_1, D^*)$, visual absolute magnitude $M_{\rm V}(\lambda_1, D^*)$, and bolometric absolute magnitude $M_{\rm bol}$ \citep{DivanZorec1982,Zorec1986,ZorecBriot1991,Zorecetal2009}. A description of these calibrations can also be found in \citet{Zorecetal2009} and \citet{Aidelman2012}. \par

Apart from the quantities derived from the BCD classification, we also refer to the apparent fundamental parameters as those used to represent spectra or whatever other observed quantity emitted in a given spectral range, most frequently in the visual spectral region, by rotationally deformed stellar atmospheres using classical plane-parallel model atmospheres in radiative and hydrostatic equilibrium \citep{Hubeny1995,Castelli2003}. The issued quantities ($T_{\rm eff},\log g)$ from the use of such model atmospheres are then called apparent fundamental parameters. \par

The method to determine apparent fundamental parameters based on the BCD ($\lambda_1, D^*$) parameters can be simply applied to B stars, both normal and peculiar, because the spectral characteristics due to the photosphere and circumstellar environments are located at different wavelengths. Thanks to this separation of spectral signatures of very different origins, it is obvious that for objects in which the radiation flux is strongly modified by the gaseous or dusty circumstellar material, like in Be and B[e] stars, the BCD system leads to a more reliable determination of ($T_{\rm eff}$, $\log g$, $M_{\rm V}$, $M_{\rm bol}$) than classic models of stellar atmospheres that cannot avoid the perturbation of line spectra or spectral energy distributions (SEDs) by the emissions and absorptions produced by extended envelopes \citep{Cidaleetal2001,Zorecetal2005}. Up to now, the BCD classification system has not been systematically used for Bn stars. The measurement uncertainties affecting $T_{\rm eff}$, $M_{\rm V}$, and $M_{\rm bol}$ depend on the measurement errors made on the BCD quantities ($\lambda_1, D^*$), which on average produce $\Delta T_{\rm eff}\!\sim\pm500$ K for the late B-type stars and $\pm1500$ K for the early B-types ($T_{\rm eff}\,>$ 20000 K), $\Delta\log g\!\sim\!\pm0.2$ dex, and $\Delta M\!\sim\pm0.3$ mag both for $M_{\rm V}$ and $M_{\rm bol}$. \par

In some cases, rather significant deviations can be noted between determinations of $M_{\rm V}$ by different methods. Since it is frequently difficult to explain their origin, in this work we adopted for $M_{\rm V}$ the average of values estimated with the following different methods: the $M_{\rm V}(\lambda_1, D^*)$ BCD calibration, absolute magnitudes determined with the Hipparcos parallaxes \citep{vanLeeuwen2007}, and the calibration by \citet{Deutschman1976} of $M_{\rm V}$ as a function of MK spectral type, where the entry spectral type is the BCD spectral type/luminosity class. The $E(B-V)$ color excess that we employed to determine the absolute magnitude with parallaxes is also an average of several estimates determined through the BCD color gradient $\Phi_{\rm rb}$ \citep{Aidelman2012,Gkouvelis2016}, the color excess determined using the absorption depression at 2200~\AA\,in the far-ultraviolet (far-UV)\citep{Briot1987,ZorecBriot1991}, the $(B-V)$ intrinsic colors given by \citet{Deutschman1976} as function of the spectral type, and the $E(B-V)$ derived from the interstellar absorption curve as a function of the distance in the direction of the studied objects \citep{Neckel1980}. \par

\begin{figure}
\centering
 \includegraphics[scale=0.96]{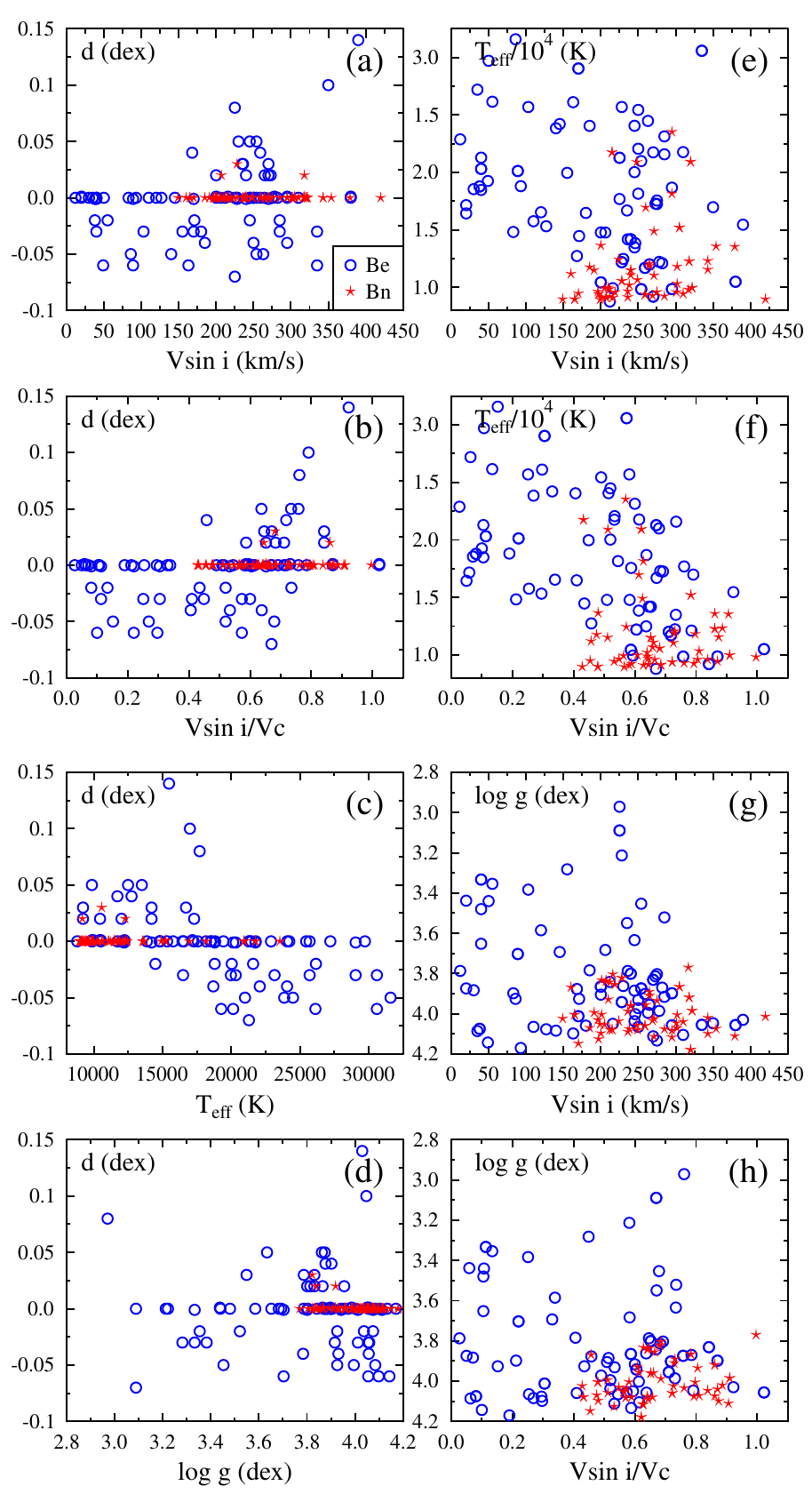} 
\caption{\label{fig_1} Graphical presentation of observed ($d, V\!\sin i, V\!\sin i/V_{\rm c}$) and apparent fundamental parameters ($T_{\rm eff}$, $\log g$) through some relations between these parameters. The blue empty circles represent Be stars, while the red stars represent Bn stars.}
\end{figure}

Although the $M_{\rm V}$ derived from parallaxes can be a good determination of the absolute magnitude for stars in which the CE does not strongly mar the visible continuum energy distribution as in most Bn stars, it can, however, be affected by such flux excesses in Be stars with detectable CE spectral signatures \citep{Ballereau1995,Moujtahid1998,Moujtahid1999}. In our Be star sample, this effect is expected to result in an uncertainty $\delta M_V\simeq0.1$~mag in the average $M_{\rm V}$. However, according to outbursts observed in Be stars, these magnitude changes can be as high as 0.3~mag \citep{Hubert1998}. \par 

The bolometric absolute magnitude $M_{\rm bol}$ adopted in this work, from which we determine the apparent bolometric luminosity $L/L_{\odot}$ of stars, was calculated with
 
\begin{equation}
\displaystyle M_{\rm bol} = M_{\rm V}+BC(T_{\rm eff}(\lambda_1, D^*))
,\end{equation}  

\noindent where $BC(T_{\rm eff})$ is the bolometric correction applied by taking into account the warnings put forward by \citet{Torres2010}. \par

We had no high-resolution blue spectra to determine reliable $\log g$ parameters by fitting models of stellar atmospheres. This quantity is  very uncertain  since estimates based on models of stellar atmospheres and evolutionary tracks do not lead to the same value \citep{Gerbaldi1993,Aidelman2012}. Therefore, we adopted an indirect estimate using models of stellar evolution without rotation \citep{Ekstrom2012} by interpolating in the evolutionary tracks the stellar $M/M_{\odot}$ as a function of $L/L_{\odot}$, determined as mentioned above, and the $T_{\rm eff}(\lambda_1, D^*)$. \par

In Tables~\ref{tabla1_Be} and \ref{tabla1_Bn} are reported the fundamental parameters derived as described above using the BCD spectrophotometric quantities together with other calibrations and methods. The $E(B-V)$ values are listed in Col. 6, while $M_{\rm V}$, $T_{\rm eff}$ and $\log~L/L_{\odot}$, together with their uncertainties, are shown in Cols. 7 to 12. From the effective temperature $T_{\rm eff}(\lambda_1, D^*)$ and the bolometric luminosity $L/L_{\odot}$, we estimated another series of derived parameters using models of stellar evolution \citep{Ekstrom2012}. Tables~\ref{tabla2_Be} and \ref{tabla2_Bn} list these values for Be and Bn stars, respectively: stellar mass $M/M_{\odot}$ and its uncertainty are given in Cols. 2 and 3; values for $\log g$ and $\sigma_{\log g}$ are given in Col. 4 and 5; the stellar radius $R/R_{\odot}$ and its uncertainty are given in Cols. 6-7; equatorial critical velocity $V_{\rm c}$, together with its uncertainties in Col. 8 and 9; and the stellar age $t$ in terms of the fractional age $t/t_{\rm MS}$, where $t_{\rm MS}$ is the time spent by a star of a given mass in the MS evolutionary phase, is given in Col. 10, and its uncertainties are listed in Col. 11. \par

For a quick overview of all these quantities, Fig.~\ref{fig_1} gives a graphical presentation of observed ($d, V\!\sin i, V\!\sin i/V_{\rm c}$) and apparent fundamental parameters ($T_{\rm eff}$, $\log g$) through some relations between them. In this figure we note that B stars acquire their classification as Bn stars for apparent $V\!\sin i \gtrsim 50$ km~s$^{-1}$ (Fig.~\ref{fig_1}a) or $V\!\sin i/V_{\rm c} \gtrsim 0.4$ (Fig.~\ref{fig_1}b) and $T_{\rm eff} \lesssim 22500$ K (Fig.~\ref{fig_1}e,f). Also, the Bn stars in our sample have $\log g  \gtrsim 3.8$, while Be stars may attain $\log g\sim3.0$ (Fig.~\ref{fig_1}g,h). \par

\begin{figure}
\centering
 \includegraphics[scale=0.96]{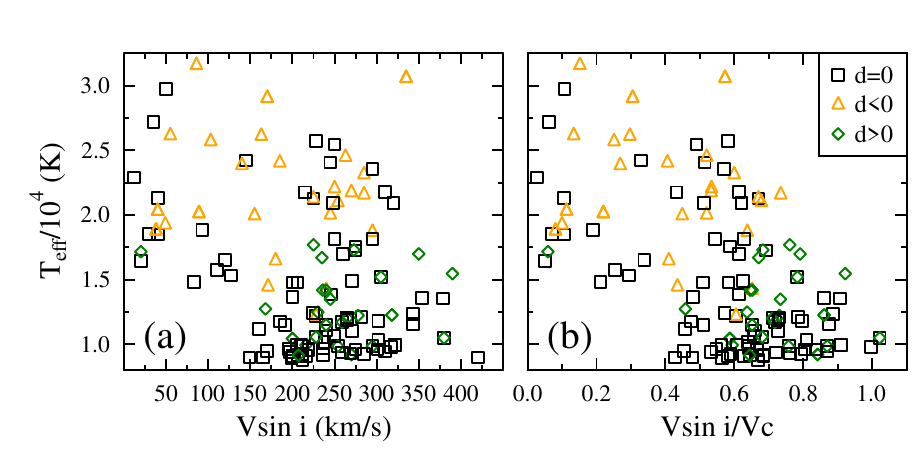} 
\caption{\label{Teff_Vsini} Apparent $V\!\sin i$ (a) and $V\!\sin i/V_{\rm c}$ (b) parameters for Be and Bn stars against $T_{\rm eff}$. Different symbols represent stars with $d<0$ (orange triangles), $d>0$ (green diamonds), and $d=0$ (black squares).}
\end{figure}

In Fig. \ref{Teff_Vsini} we show the $T_{\rm eff}$ values against $V\!\sin i$ (a) and $V\!\sin i/V_{\rm c}$ (b), where the different symbols represent the appearance of the sBD for Be and Bn stars. Black squares indicate stars without a sBD, while orange triangles and green diamonds represent stars with the sBD in emission and absorption, respectively. It can be seen that the sBD in absorption appears for stars with $T_{\rm eff} \lesssim 22500$ and $V\!\sin i \gtrsim 250$ km\,s$^{-1}$, i.e., star-CE systems likely seen equator-on. On the other hand, the sBD in emission appears however likely for stars with $T_{\rm eff} \gtrsim 15000$ K and values of $V\!\sin i \lesssim 250$ km\,s$^{-1}$, which indicates that low inclination angles are privileged. \par

\subsection{Relations between emission intensities and fundamental parameters}\label{rbeiafp} 

In Sect.~\ref{apppar} we determined the occurrence of the sBD as a function of $T_{\rm eff}$ and $V\!\sin i$ (or $V\!\sin i/V_{\rm c}$) for Be and Bn stars. We can go even a little further by searching for relations between the intensity of different emission (or absorption) signatures due to the CE and some fundamental parameters for Be stars. Since the formation regions of the emission in the H$\alpha$ line and that of the sBD component are not the same, we expect to get information that may have an impact on the modeling of the physical structure of the CEs and perhaps of their formation process. \par

The amount of emission in the H$\alpha$ line is presented in terms of its total equivalent width $W$ in~\AA\,, where positive values of $W$ correspond to emission lines and the flux $F_{\rm H\alpha}$ (erg/cm$^2$/sec). The total equivalent width is given as

\begin{equation}
\displaystyle W = W_{\rm e}+W_{\rm ph},
\end{equation}

\noindent where $W_{\rm e}$ represents the equivalent width of the emission superimposed to the underlying photospheric absorption line profile and $W_{\rm ph}$ is the equivalent width either of the total photospheric component or a fraction that was fitted with an empirical relation introduced by \citet{Ballereau1995}, whose efficiency was since then proved by several authors \citep{Chauville2001,Levenhagen2006,Arias2018}. The flux in the H$\alpha$ line emission is calculated as $F_{\rm H\alpha}$ = $W\times F^c_{\rm H\alpha}$, where $F^c_{\rm H\alpha}$ is the continuum flux at H$\alpha$ line for the apparent ($T_{\rm eff},\log g$) stellar parameters. In Table~\ref{tabla_flux} we present in turn values of $W_{\rm e}$, $W$, $F_{\rm H\alpha}$, and $d$. \par

\begin{figure}
\centering\includegraphics[scale=0.96]{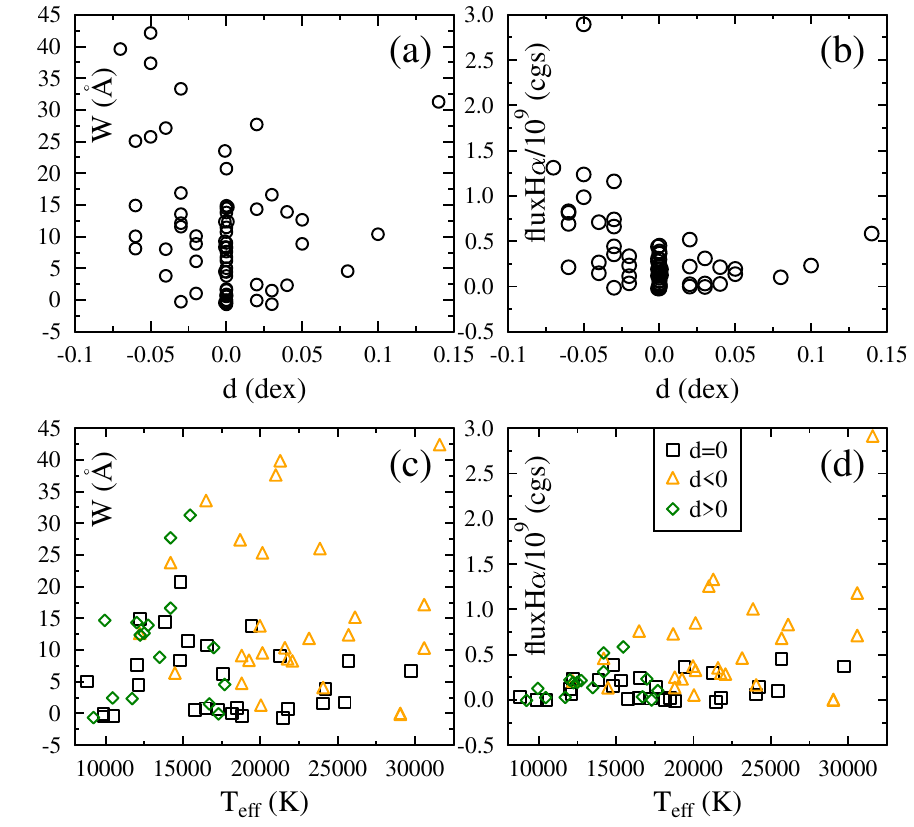} 
\caption{\label{fig_1n} (a): Total equivalent widths in~\AA\,of the H$\alpha$ line emission components for Be stars against the sBD component $d$ in dex. (b): Fluxes of the H$\alpha$ line emission components against the sBD $d$ in dex. (c): Total equivalent widths in~\AA\,of the H$\alpha$ line emission components against the apparent effective temperature. (d): Fluxes of the H$\alpha$ line emission components against the apparent effective temperature.} 
\end{figure}

The results are shown in Fig.~\ref{fig_1n} where only tendencies are apparent but not tight correlations. A strong tendency is noted in Figs.~\ref{fig_1n}a and b for enhanced sBD in emission ($d<0$) as the emission in the H$\alpha$ line increases, but the spread of points is very large, which indicates that probably different physical and geometrical structures and aspect angles of the CE can produce a given value of $d$ (positive, negative, or even $d=0$) for different amounts of emission in H$\alpha$ line and vice versa. In Fig.~\ref{fig_1n}c we observe that Be stars without a sBD generally present small intensities in the H$\alpha$ line over all the $T_{\rm eff}$ values. Also, for $T_{\rm eff} \gtrsim 1500 0$, where the sBD appears mostly in emission, the intensity of the H$\alpha$ line become stronger as Teff increases. For $T_{\rm eff} \lesssim 17000$ and a sBD in absorption there is no clear effect of the effective temperature onto the strength of $d>0$. For fluxes $0 \lesssim F_{\rm H\alpha}/10^9 \lesssim 0.5$ (erg/cm$^2$/sec), it is possible to have $d=0$, but also all possible values of $d<0$ or $d>0$. Taking into account that early Be-type stars present a greater degree of variability \citep{Labadie2017}, we expect to observe a wider range of H$\alpha$ line intensity for higher $T_{\rm eff}$. Thus, the measurement of the intensity of the H$\alpha$ line obtained from spectra taken at  different dates than those of our low-resolution spectra may not represent the same state of the disk. \par 

\subsection{Parent nonrotating counterpart parameters}\label{pnrcpar}

In what follows we study the properties of our program Be and Bn stars using stellar fundamental parameters corrected for rotational effects. The apparent fundamental parameters inferred directly from the observed spectra of stars are mainly functions of the stellar mass $M/M_{\odot}$, fractional age $t/t_{\rm MS}$, inclination angle $i$ of the rotation axis, and the law representing the angular velocity distribution inside and on the stellar surface $\Omega(r,\theta)$ corresponding to the current stellar evolutionary stage ($r$ is the radial distance from the stellar center and $\theta$ is the co-latitude angle). This stellar angular velocity resumes the whole history of the angular momentum redistribution processes and the losses through the stellar mass-loss phenomena the star underwent since the zero age main sequence (ZAMS) and even before during the pre-MS phase. We do not have any information on the internal distribution of the stellar angular velocity. Therefore, to interpret the apparent rotation parameter $V\!\sin i$ at the current stellar evolutionary stage and the remaining fundamental parameters, we have to assume some rotation law. At the moment, our only option is to assume $\Omega(r,\theta)_{\rm surface}\!=\!\Omega_o$, i.e., $\Omega$ is uniform over the stellar surface (shellular rotation). In the present approach we also assume that $\Omega_o$ is a function of time, $\Omega_o=\Omega_o(t)$, and that it depends on the internal angular velocity evolution as described by the models calculated by \citet{Maeder2000}, \citet{Ekstrom2008} and \citet{Ekstrom2012}. To take into account the effects of the rotation on the observed spectral characteristics, according to these models and assumptions, the following formal system of equations has to be solved in terms of the parameters $M/M_{\odot}$, $t/t_{\rm MS}$, $\Omega/\Omega_{\rm c}$ and $i$:

\begin{equation}
\begin{array}{rcl}
\displaystyle T_{\rm eff}^{\rm app} & = & \displaystyle T_{\rm eff}^{\rm pnrc}(M,t)\ C_{\rm T}(M,t,\eta,i)\\
\\
\displaystyle g_{\rm eff}^{\rm app} & = & \displaystyle g_{\rm eff}^{\rm pnrc}(M,t)\ C_{\rm G}(M,t,\eta,i)\\
\\
\displaystyle L^{\rm app} & = & \displaystyle L^{\rm pnrc}(M,t)\ C_{\rm L}(M,t,\eta,i)\\
\\
\displaystyle \frac{(V\!\sin i)_{\rm app}}{V_{\rm c}(M,t)} & = & \displaystyle \left[\frac{\eta}{R_{\rm e}(M,t,\eta)/R_{\rm c}(M,t)}\right]^{1/2}\!\!\!\!\sin i-\frac{\Sigma(M,t,\eta,i)}{V_{\rm c}(M,t)}, 
\label{eq_8}
\end{array}
\end{equation}

\noindent where $R_{\rm e}(M,t,\eta)$ and $R_{\rm c}(M,t)$ are the actual and the critical stellar equatorial radii corresponding to the current stellar evolutionary stage, respectively, which are determined using our 2D models of rigidly rotating stars \citep{Zorec2011,Zorec2012}. The quantity $\eta=(\Omega/\Omega_{\rm c})^2[R_{\rm e}/R_{\rm c}]^3$ is the ratio of the centrifugal to the gravitational acceleration in the equator, and approaches 1 more slowly than $\Omega/\Omega_{\rm c}$ and $V/V_{\rm c}$ when the stellar rotation becomes critical. The left-hand side of Eq.~\eqref{eq_8} is associated with the apparent fundamental parameters determined in Sect.~\ref{apppar}. On the right side of Eq.~\eqref{eq_8}, $T_{\rm eff}^{\rm pnrc}(M,t)$, $g_{\rm eff}^{\rm pnrc}(M,t)$, $L^{\rm pnrc}(M,t)$ are the parent nonrotating counterpart (pnrc) parameters effective temperature, surface gravity, and bolometric luminosity \citep[see definition of pnrc in][]{Fremat2005}. The functions $C_{\rm T}(M,t,\eta,i)$, $C_{\rm G}(M,t,\eta,i),$ and $C_{\rm L}(M,t,\eta,i)$ carry information relative to the change of parameters due to the geometrical deformation of the rotating star and of its  GD over the observed hemisphere. Stoeckley's correction $\Sigma(\eta,i,M.t)$ of the $V\!\sin i$ parameter for GD depends on a power exponent $\beta_1$ of the effective temperature, which depends on the colatitude angle $\theta$ and the surface angular velocity ratio $\Omega/\Omega_{\rm c}$ \citep{Zorec2016,Zorec2017b}. Since we do not master these dependencies, we take as an approximation $\beta_1=1$, which produces an overestimation of the GD effect. The method used to solve the system of equations in Eq.~\eqref{eq_8} is detailed in \citet{Zorec2016}. \par

An entirely consistent determination of the pnrc parameters in Eq.~\eqref{eq_8} demands that the entry apparent quantities $(T_{\rm eff},\log g_{\rm eff},L/L_{\odot})$ be independent. This was attempted by \citet{Zorec2016} for a sample of bright Be stars for which all the required observational material was available along with information on the photometric variation of the studied objects useful to correct the SEDs from the perturbations introduced by the CE. On the contrary, in the present case we do not have this information for all the program stars. However, to treat the entire sample in the same way we attempted to obtain a first order estimate of the rotational effects on the fundamental parameters by assuming that they are rapid rotators having the same surface angular velocity ratio $\Omega/\Omega_{\rm c}=0.95$. The pnrc parameters are meant to represent the studied objects as they were without rotation. These are given in Tables~\ref{tabla3_Be} and \ref{tabla3_Bn} for Be and Bn stars, respectively. Columns 2 to 7 present $T_{\rm eff}$, $\sigma_{T_{\rm eff}}$, $\log L/L_{\odot}$, $\sigma_{\log L/L_{\odot}}$, $\log g$ and $\sigma_{\log g}$, Col. 8 gives $M/M_{\odot}$ with its uncertainties in Col. 9, $V\!\sin i$ and $V_{\rm c}$ with their uncertainties are presented in Cols. 10-13, while $t/t_{\rm MS}$ and $\sigma_{t/t_{\rm MS}}$ are given in Cols. 14 and 15. \par

\begin{figure}
\centerline{\includegraphics[scale=0.55]{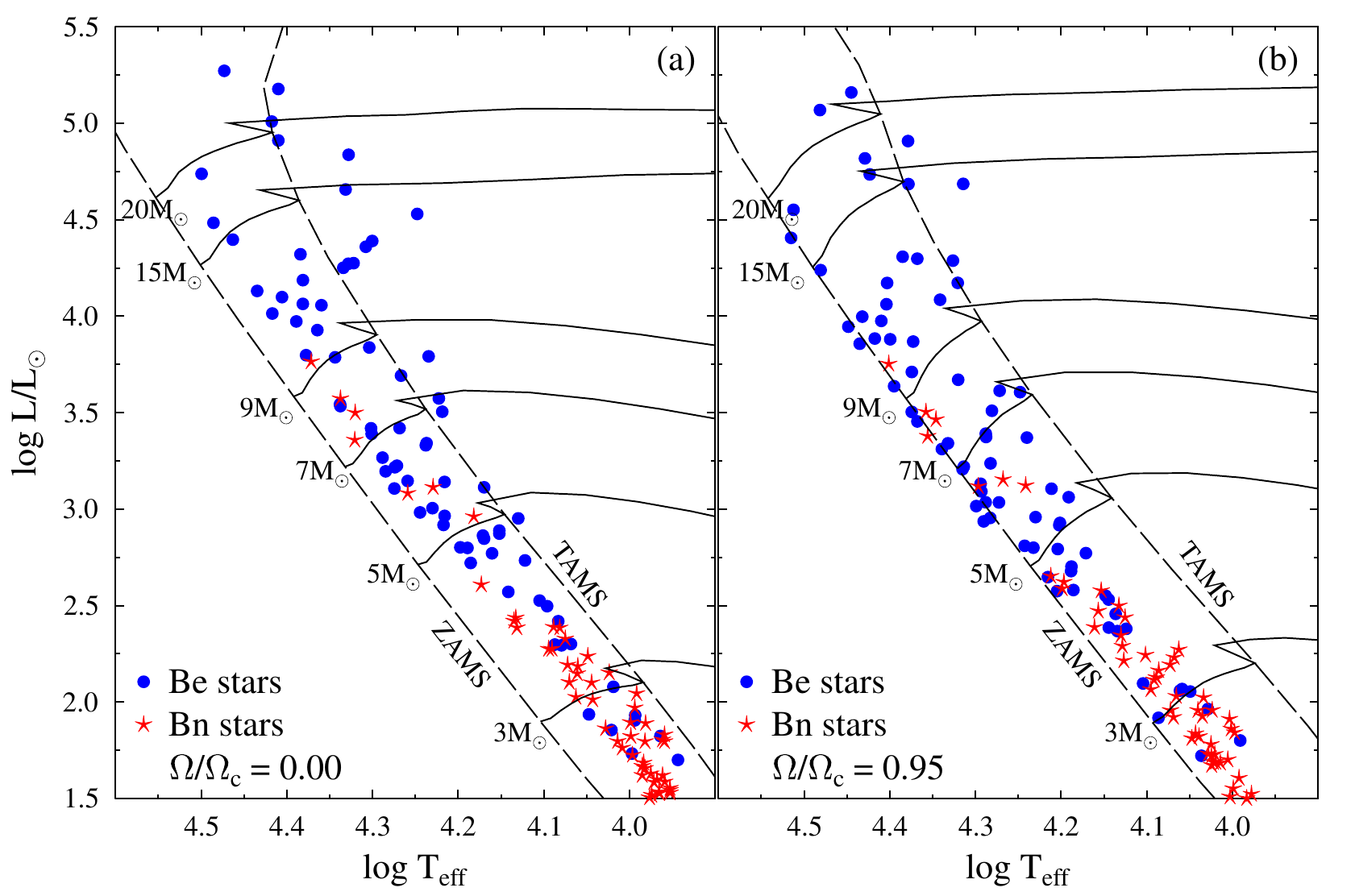}}
\caption{\label{hr_lt}(a) Hertzsprung-Russell (HR) diagram of the studied Be (blue points) and Bn stars (red stars) obtained with apparent ($\log L/L_{\odot},\log T_{\rm eff}$) parameters (not treated for rotational effects). (b) HR diagram drawn with parameters corrected for rotational effects, assuming that all stars rotate at $\Omega/\Omega_{\rm c}=0.95$. The evolutionary tracks are from \citet{Ekstrom2012} without rotation (a) and with rotation, with $V_o=300$ km\,s$^{-1}$ in the ZAMS (b).}
\end{figure}

\section{Evolutionary status of Be and Bn stars}\label{esobbs}
\subsection{Hertzsprung-Russell diagram of our program stars}\label{hrbebn}

Since Be and Bn stars are both rapid rotators, we can ask whether they have similar structures or other common properties that would allow us to consider both of these stars as members of a single population and, in particular, to think of Bn stars as potential Be stars. We begin by comparing their evolutionary state by searching for possible signatures using the observed (apparent) parameters and those corrected for rotational effects. Figure~\ref{hr_lt}a shows the HR diagram of the studied objects obtained with the apparent ($\log L/L_{\odot},\log T_{\rm eff}$) parameters. Figure~\ref{hr_lt}b shows the same HR diagram, but drawn with pairs ($\log L/L_{\odot},\log T_{\rm eff})_{\Omega/\Omega_{\rm c}}$ calculated under the assumption of $\Omega/\Omega_{\rm c}=0.95$. The evolutionary tracks are from \citet{Ekstrom2012} for $\Omega/\Omega_{\rm c}=0$ in Fig.~\ref{hr_lt}a and for $\Omega/\Omega_{\rm c}\neq 0$ in Fig.~\ref{hr_lt}b, where the adopted equatorial linear velocity in the ZAMS is $V_o=300$ km~s$^{-1}$. \par

The main difference seen in the HR diagrams of Fig.~\ref{hr_lt} is that Bn stars have masses $M \lesssim 9M_{\odot}$, while Be stars range from $3M_{\odot} \lesssim M $ to $M \lesssim 20M_{\odot}$. The correction of bolometric luminosities and effective temperatures for rotation effects brings back the objects to younger evolutionary stages in the  MS and transforms apparent blue supergiants into MS objects. Another characteristic that comes from the HR diagram is that our Bn stars are close to the ZAMS, while in the same mass interval ($4 \lesssim M/M_{\odot} \lesssim 9$) Be stars occupy the entire MS evolutionary span. Bn stars are more numerous than Be stars for masses $M/M_{\odot} \lesssim 4$, but both roughly share the same evolutionary domain. \par

\subsection{Distribution of stellar ages in our sample}\label{dabebn}
 
\begin{figure}
\centerline{\includegraphics[scale=0.75]{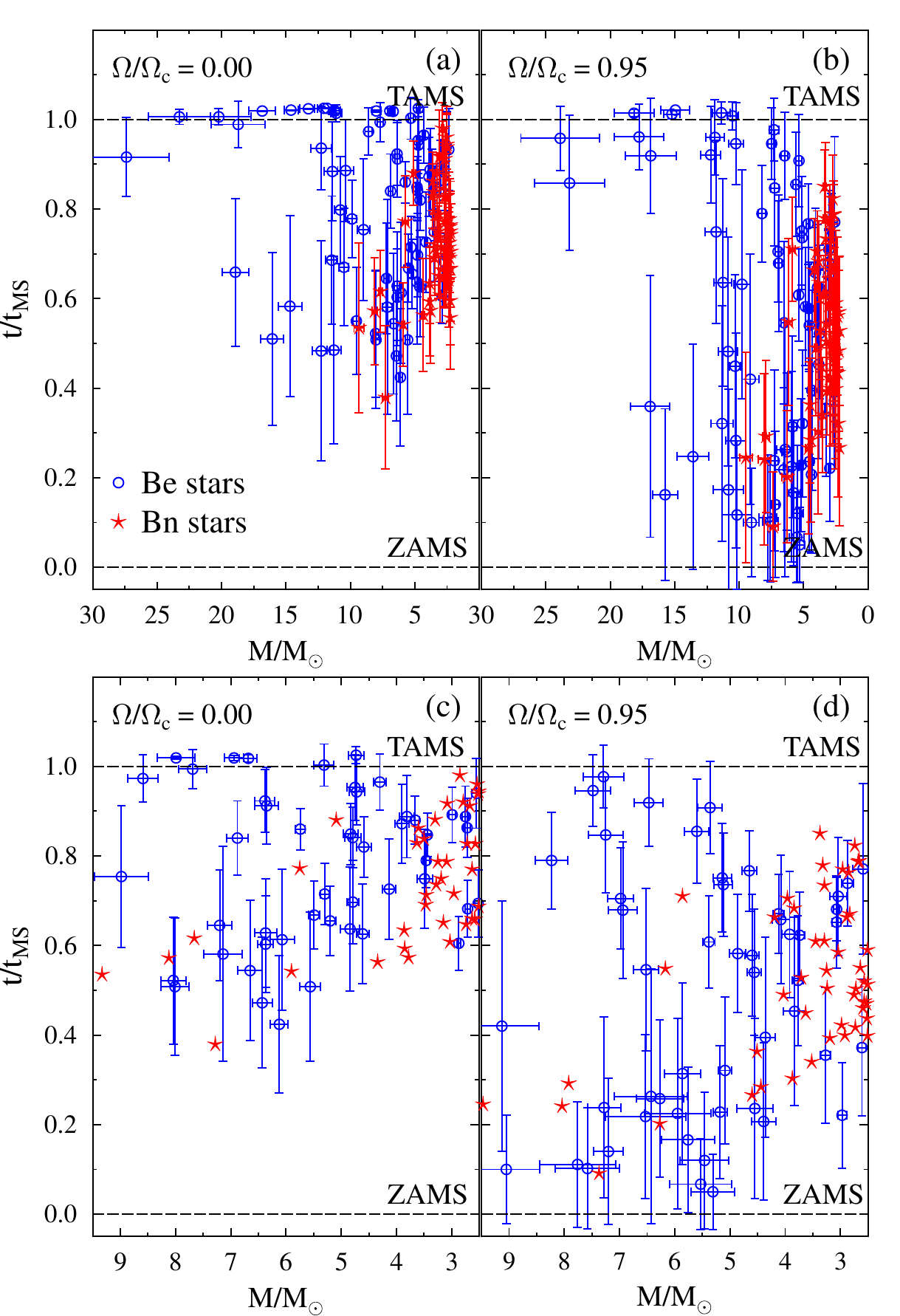}} 
\caption{\label{hr_mt} Age vs. mass diagrams using the apparent pairs ($M/M_{\odot},t/t_{\rm MS}$) not treated for rotation [panels (a) and (c)], and using parameters corrected for rotational effects [panels (b) and (d)]. Panels (c) and (d) are zooms of (a) and (b) for masses $M/M_{\odot}< 10$, respectively. Error bars indicate uncertainties affecting the $t/t_{\rm MS}$ and $M/M_{\odot}$ determinations.}
\end{figure}

Another way to describe the evolutionary stage of the stu\-died objects is using masses and ages determined from Eq.~\eqref{eq_8} with the help of evolutionary tracks without or with rotation as previously done in \citet{Zorecetal2005} and \citet{Martayan2007}. From the inferred age ratios $t/t_{\rm MS}$ and masses $M/M_{\odot}$, we obtain the diagrams shown in Fig.~\ref{hr_mt}a,b,c,d. In Figs.~\ref{hr_mt}a and b the apparent parameters ($M/M_{\odot},t/t_{\rm MS}$) are plotted without and with correction for rotation effects, respectively. Figures~\ref{hr_mt}c and d depict zooms of Figures~\ref{hr_mt}a and b for masses $M/M_{\odot}< 10$, respectively, to better separate the behavior of Bn stars (red stars) from that of Be stars (blue points). The ($M/M_{\odot},t/t_{\rm MS}$) diagrams suggest that Bn stars seem to fill up an apparent gap of Be stars nested in the mass range $2 \lesssim M/M_{\odot} \lesssim 4$, which encompasses a more or less large evolutionary span. The solution pairs ($M/M_{\odot},t/t_{\rm MS}$) for our program objects locate both Be and Bn stars in the second half of the MS when rotational effects are not accounted for, while under the assumption of $\Omega/\Omega_{\rm c}=0.95$ for all objects, the points for both type of stars scatter from the ZAMS to the terminal age main sequence (TAMS). \par

If Eq.~\eqref{eq_8} is solved assuming that the input parameters are entirely independent, we obtain a different estimate of $\Omega/\Omega_{\rm c}$ for each star. However, the new diagram ($M/M_{\odot},t/t_{\rm MS}$) thus obtained does not differ strongly from Fig.~\ref{hr_mt}. Nevertheless, even with the program star data available to us, it is impossible for us to derive a set of entirely independent input parameters $(T^{\rm app}_{\rm eff},g^{\rm app}_{\rm eff},L^{\rm app},V\!\sin i_{\rm app})$. We can think of the group of three parameters $T^{\rm app}_{\rm eff}$, $g^{\rm app}_{\rm eff}$ (or $L_{\rm app}$), and $V\!\sin i_{\rm app}$ as independent, but $g^{\rm app}_{\rm eff}$ and $L_{\rm app}$ are mutually dependent because these parameters are obtained from $g^{\rm app}_{\rm eff}=g[T^{\rm app}_{\rm eff},L_{\rm app}]$ or $L_{\rm app}=L_{\rm app}[T^{\rm app}_{\rm eff},g^{\rm app}_{\rm eff}]$. For this reason, our results are only considered to be a first attempt to determine the evolutionary status of the program stars as rapid rotators. \par

In the mass interval $M/M_{\odot} \gtrsim 5$ it is likely that \mbox{$(t/t_{\rm MS})_{\rm Bn} \lesssim (t/t_{\rm MS})_{\rm Be}$}. In Figure~\ref{hr_mt}a, the higher values of $t/t_{\rm MS}$ ratios of Bn stars with $M/M_{\odot} \lesssim 4$ seem to suggest that they will never attain the required conditions to display the Be phenomenon during the MS evolutionary phase. Contrarily, in the diagrams where the parameters were corrected for rotational effects, we see that Bn stars can have age ratios $0.2~\lesssim~(t/t_{\rm MS})_{\rm Bn}~\lesssim~0.8$ if $M/M_{\odot} \lesssim 4$, whereas they have $0.0 \lesssim (t/t_{\rm MS})_{\rm Bn} \lesssim 0.2$ for masses $M/M_{\odot} \gtrsim 5$. According to \citet{Aidelman2018}, the Be phenomenon is observed along the whole MS, and appears on average at an earlier age in massive stars than in the less massive stars. Thus, the possibility that the most massive Bn stars display the Be phenomenon at any time should not be excluded. \par

\begin{figure*}[t!]
\centerline{\includegraphics[scale=0.64]{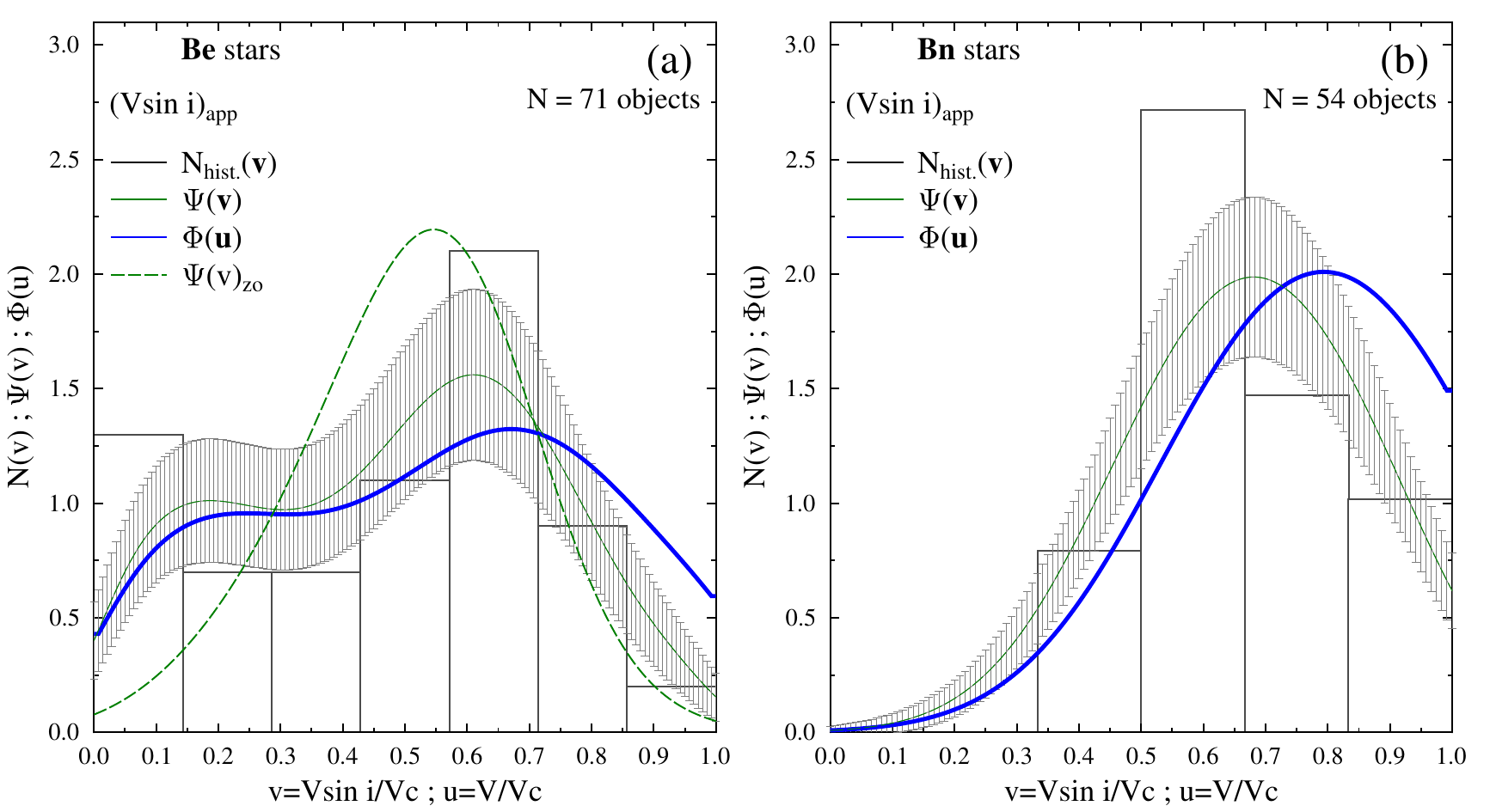}}
\centerline{\includegraphics[scale=0.64]{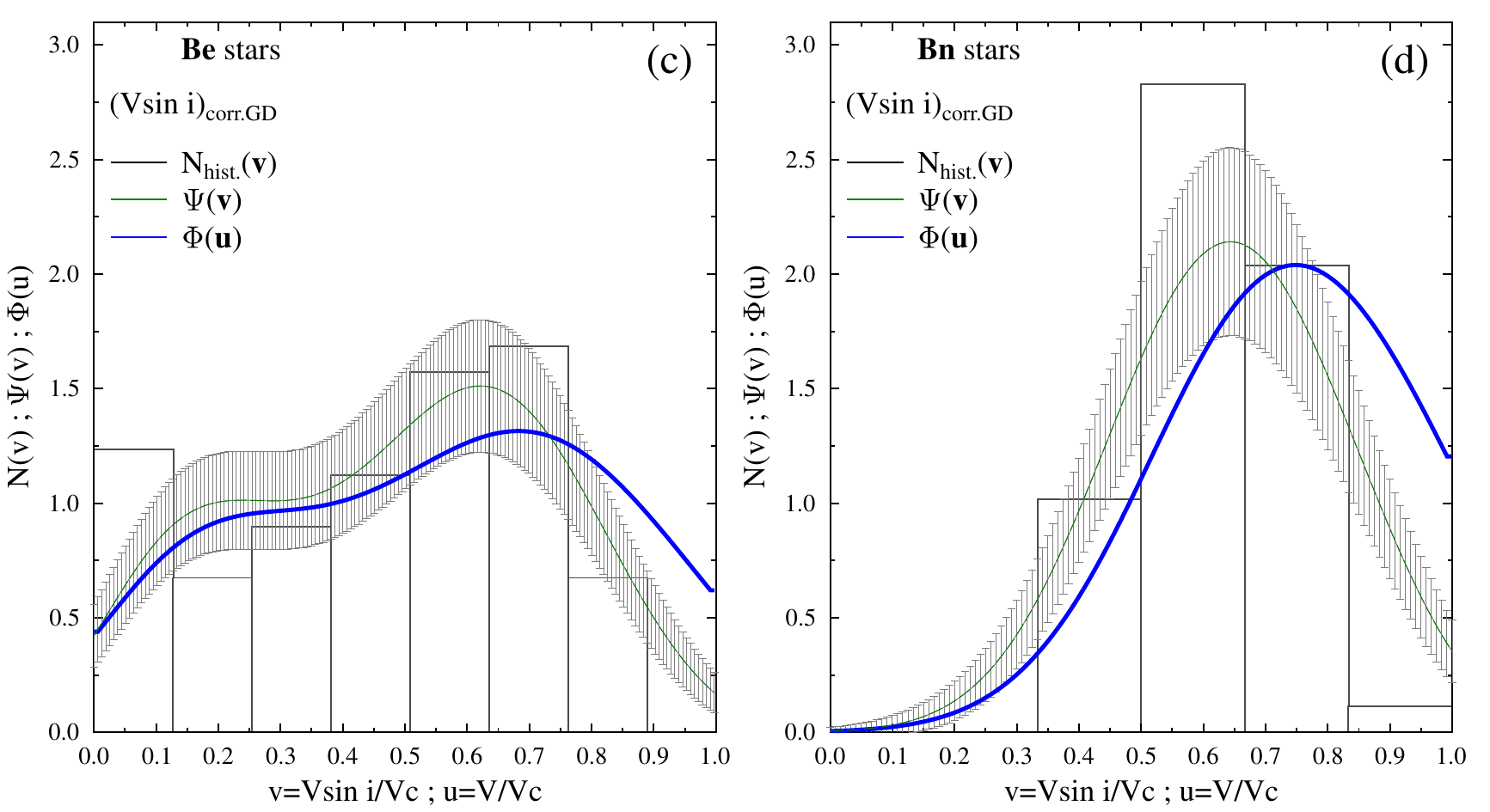}}
\centerline{\includegraphics[scale=0.64]{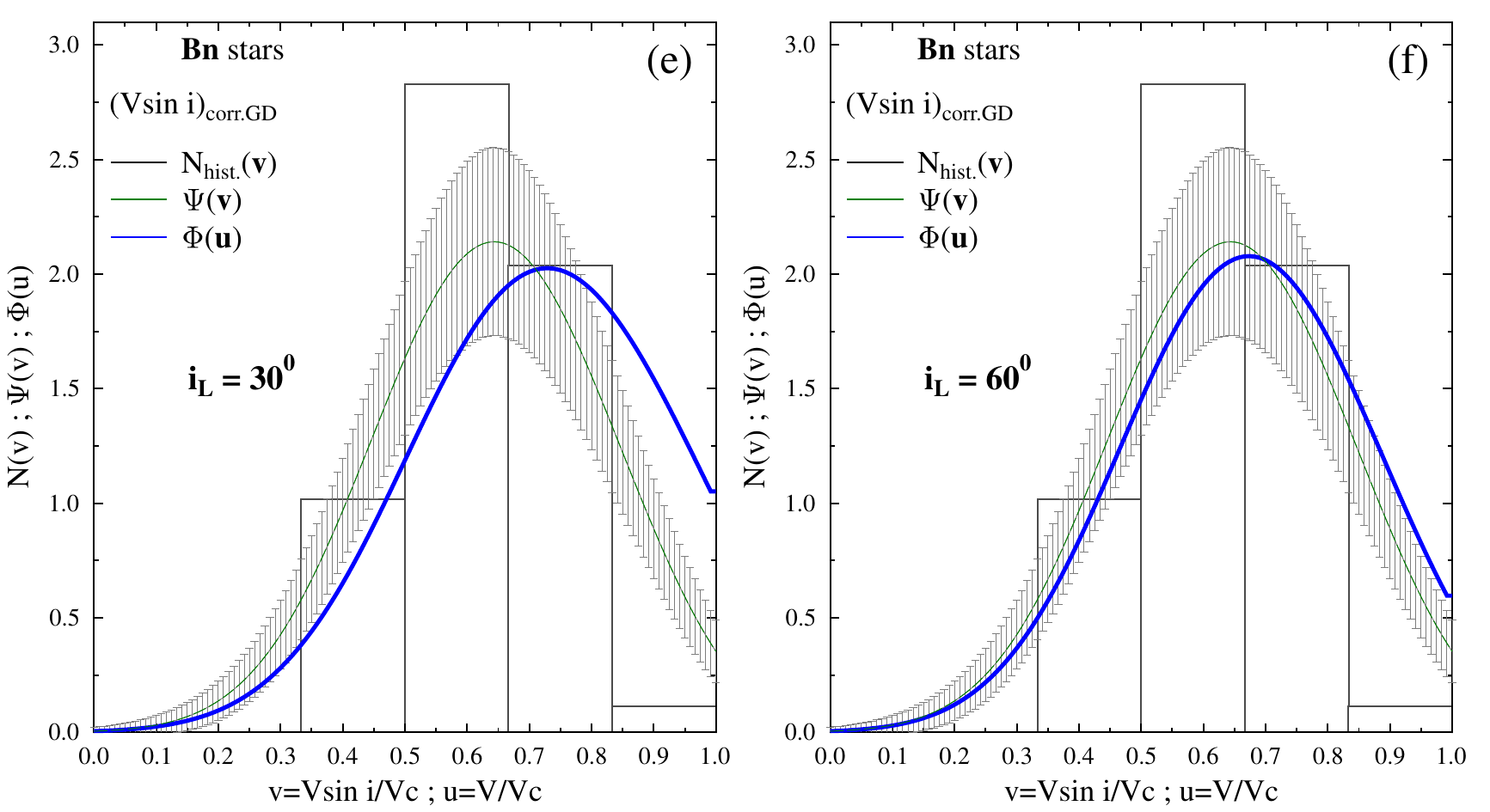}}
\caption{\label{app_vsini} Distributions without correction for GD effect. (a) Be and (b) Bn stars: Histograms of apparent velocity ratios $v=V\!\sin i/V_{\rm c}$; functions $\Psi(v)$ representing the smoothed histograms after correction for observational uncertainties (green curves); functions $\Phi(u)$ representing distributions of true velocity ratios $u=V/V_{\rm c}$ (blue curves); in (a) $\Psi(v)$ from \citet{Zorec2016} (green dashed curve). Error bars in the smoothed histograms represent the statistical uncertainties that also concern the remaining distribution. (c) Be and (d) Bn stars: Same as (a) and (b), but distributions of velocities corrected for GD effect. (e) and (f): Histograms, functions $\Psi(v)$ from (d) and distributions of true velocity ratios $u=V/V_{\rm c}$ for Bn star (blue curves), corrected for GD effect and for density probabilities of inclination angles restricted at inclinations $i_{\rm L}=30\degr$ and $i_{\rm L}=60\degr$, respectively.}
\end{figure*}

Regarding Bn stars, two questions can then be raised: First, do they  have the required properties as rotators to display the Be phenomenon before they attain the TAMS? Second,  are there CEs formed around these stars whose existence has not yet been detected because their temperatures are too low to excite emission lines? \par 

\section{Distribution of rotational velocities at the current stellar evolutionary stage}\label{davbebn}

It is worth noting that there are selection effects marring the Be and Bn stellar sets employed in this work. On one hand, Be stars seen pole-on were somewhat privileged because they exhibit the most prominent sBD. On the other hand, Bn stars are currently identified as such when they are seen rather equator-on, otherwise the lack of emissions in their spectra do not enable us to identify those which rotate rapidly but are seen pole-on. \par

Figures~\ref{app_vsini}a and \ref{app_vsini}b show the distributions of the observed (apparent) $V\!\sin i$ values of the program Be and Bn stars, respectively. In these figures, histograms correspond to the raw observed apparent velocity ratios $v=V\!\sin i/V_{\rm c}$. The superimposed green curves $\Psi(v)$ describe the smoothed distributions of the ratios $v=V\!\sin i/V_{\rm c}$ corrected for observational uncertainties. We preferred to use the ratios $V\!\sin i/V_{\rm c}$ instead of the $V\!\sin i$ parameters because the critical equatorial velocity $V_{\rm c}$ is estimated consistently with the mass and evolutionary state of each star, which then minimizes somewhat mass- and evolution-related effects on the distributions \citep{Zorec2016}. The class-steps of the histograms are established according to the bin-width optimization method by \citet{Shimazaki2007}. The smoothed version of the frequency density distribution of ratios $v=V\!\sin i/V_{\rm c}$ corrected for measurement uncertainties were calculated using kernel estimators \citep{Bowman97}, where each observed parameter $v=V\!\sin i/V_{\rm c}$ is represented by a Gaussian distribution whose dispersion is given by the standard deviation of individual $V\!\sin i$ estimates. \par

In Fig.~\ref{app_vsini}a we also added the distribution $\Psi(v)$ of ratios $v=V\!\sin i/V_{\rm c}$ corrected for observational uncertainties determined in \citet{Zorec2016} for a sample nearly four times larger of Be stars than that used in this work. Comparing our $\Psi(v)$ and the same distribution for Be stars in \citet{Zorec2016} we note the effect carried by the bias affecting our Be sample, which privileges stars with as large as possible sBDs, and consequently does not warrant the random distribution of the inclination angles. Likewise, according to comments in Sect.~\ref{dabebn}, pole-on Bn stars are systematically missing in our sample, so that their distribution of apparent rotational velocity do not respect the randomness of inclination angles either. Nonetheless, we produced distributions of ratios of true velocities $u=V/V_{\rm c}$ for both our Be and Bn stars as the inclination angles were  distributed at random. We obtain thus the smoothed distributions $\Phi(u)$ shown in Figs.~\ref{app_vsini}a,b,c,d (blue curves). \par

\begin{figure}
\centerline{\includegraphics[scale=0.7]{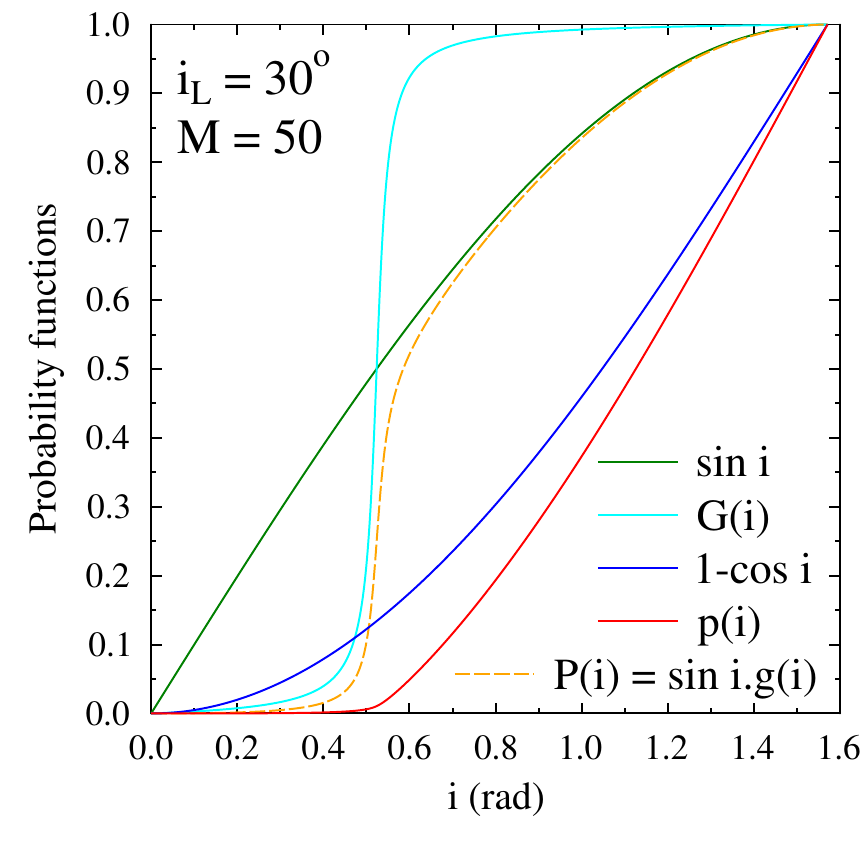}} 
\caption{\label{gill} Probability functions entering the definition of the guillotine function and obliterating the probability distribution of inclination angles.}
\end{figure}

The lack of randomness of inclination angles in the distributions of Be rotational velocities presented in this work seems impossible to correct. We can, however, attempt to account for this lack in the distributions of Bn rotational velocities when transforming the distribution of ratios $v=V\!\sin i/V_{\rm c}$ into $u=V/V_{\rm c}$ ratios of true velocity ratios. To this end, we obliterate the probability density distribution of inclination angles from $i=0^o$ up to some limiting inclination $i_{\rm L}$ using a ``guillotine''\footnote{ In this work, ``guillotine'' function means to obliterate the effectiveness of the probability function $P(i)$ in a given interval of inclination angles. We borrowed the term guillotine from the guillotine factor used to reduce Kramer's opacity in stellar interiors of low-mass stars \citep{Eddington1932}.} function $G(i)$ as follows:

\begin{equation}
\begin{array}{rcl}
\displaystyle P(i)\,di & = \displaystyle & \sin i.G(i)d\,i \\
\displaystyle G(i) & = & \displaystyle A(M,i_{\rm L})\left[\frac{g(i)-g(0)}{g(\pi/2)-g(0)}\right] \\
\displaystyle g(i) & = & \displaystyle \arctan[M(i-i_{\rm L})] \\
\displaystyle p(i) & = & \displaystyle \int_{0}^{i}\sin i.G(i)\,di
\end{array} 
,\end{equation}

\noindent where $P(i)$ is normalized, i.e., $\int_{0}^{\pi/2}P(i)\,di=1$, $A(M,i_{\rm L})$ is a normalization constant, and $M$ is the ``sharpness'' of the cut. Figure~\ref{gill} shows the behavior of functions entering the restrained probability function $p(i)$. Figures~\ref{app_vsini}e and f show the distributions of ratios $u=V/V_{\rm c}$ of true velocities obtained by imposing $M=50$, $i_{\rm L}=30\degr$ and $i_{\rm L}=60\degr$. All transformations of distributions of $v=V\!\sin i/V_{\rm c}$ into those of $u=V/V_{\rm c}$ are carried out in this work using the Richardson-Lucy deconvolution method \citep{Richardson72,Lucy1974}. We note that increasing the value of $M$ no sensitive effects are produced on the distribution functions obtained. The most outstanding results we can draw from Figs.~\ref{app_vsini}c and d are that changing the value of $i_{\rm L}$ neither the skewness of the distribution is changed nor the mode (position of the maximum). The only characteristic that seems to vary a little more concerns the number of  fastest stars: the higher the limiting inclination angle $i_{\rm L}$ the lower is the number of fast rotators, and accordingly the larger is the number of objects just behind the mode. \par

\section{Distribution of Be and Bn star rotational velocities in the ZAMS}\label{zamsdv}

\subsection{Inference of the $V/V_{\rm c}$ velocity ratios at the current stellar evolutionary stage}\label{itvvc}

\begin{figure}
\centerline{\includegraphics[scale=0.7]{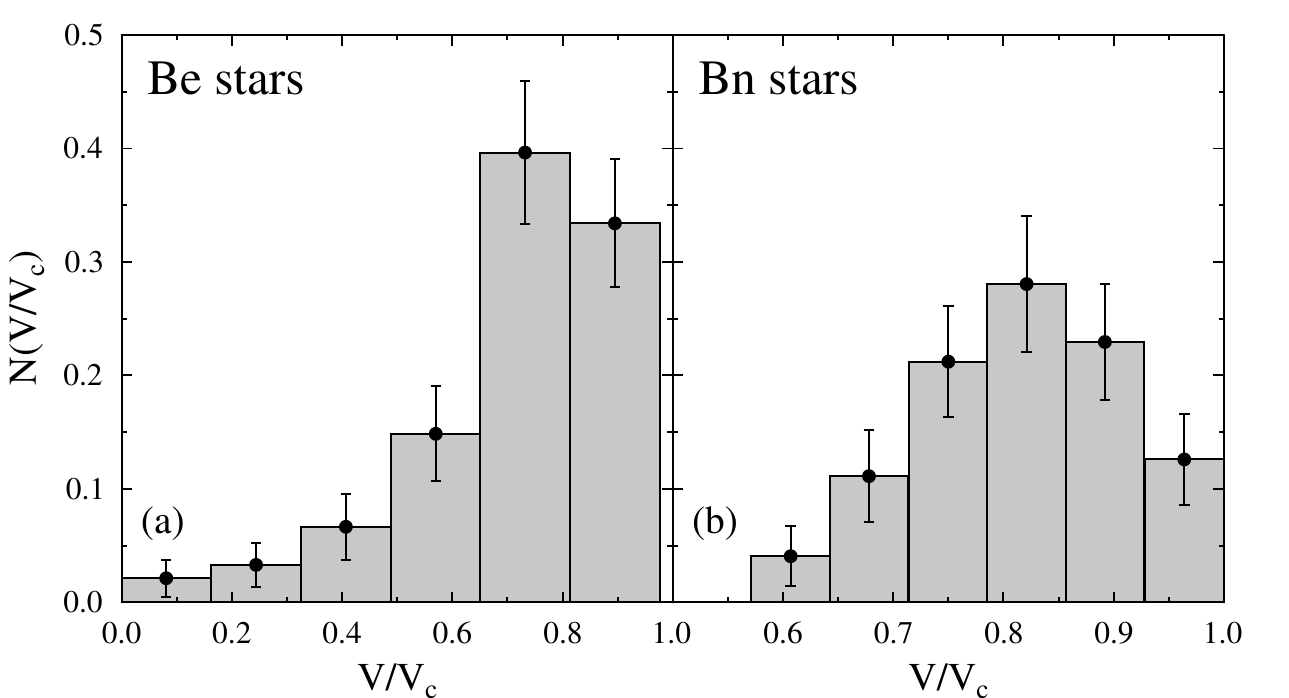}}
\caption{\label{vvc_hoy} Normalized histograms of true velocity ratios $V/V_{\rm c}$ at the current stellar evolutionary stage of Be [panel (a)] and Bn stars [panel (b)] obtained from Eq.~\eqref{eq_8}.} 
\end{figure}

There is a final comparison we can make with velocity ratios $V/V_{\rm c}$ estimated for each star individually. This can be done using the system of equations Eq.~\eqref{eq_8}, where the angular velocity ratio $\Omega/\Omega_{\rm c}$, or the ratio $V/V_{\rm c}$ of linear rotational velocities are taken into account through the parameter \mbox{$\eta=(\Omega/\Omega_{\rm c})^2(R_{\rm e}/R_{\rm c})^3=$ $(V/V_{\rm c})^2(R_{\rm e}/R_{\rm c})$}. The solution of these equations can be drawn leaving the angular velocity ratio $\Omega/\Omega_{\rm c}$ as a free parameter, or by imposing a value for it. As already commented in this paper, we have chosen the second possibility. The distribution of the $V/V_{\rm c}$ values is then necessarily different from that in Fig.~\ref{app_vsini} because no restriction is now imposed on the distribution of inclinations, while in Sect.~\ref{davbebn} we assumed randomness or limiting it to the $(i_{\rm L},\pi/2)$ interval. The results obtained for the ratios $V/V_{\rm c}$ at the current stellar evolutionary stage by solving Eq.~\eqref{eq_8} for $\Omega/\Omega_{\rm c}=0.95$ are shown in Fig.~\ref{vvc_hoy} as normalized histograms. The outstanding differences between both distributions are: First, Bn stars seem to have ratios $V/V_{\rm c}$ that are strongly concentrated to the interval $0.6 \lesssim V/V_{\rm c} \lesssim 1.0$, which is the consequence of having missed pole-on Bn stars. Second, there are very few Be stars in our sample with $V/V_{\rm c} \lesssim 0.6$. Third, Be stars have a larger proportion of objects than Bn stars with ratios approaching 0.9-1.0. Fourth, the skewness of both distributions is likely negative, i.e., the mean and median are smaller than the mode. Fifth, the mode of distributions lies around $V/V_{\rm c}\sim0.7-0.8$ for both types of objects.

\begin{figure} 
\centerline{\includegraphics[scale=0.7]{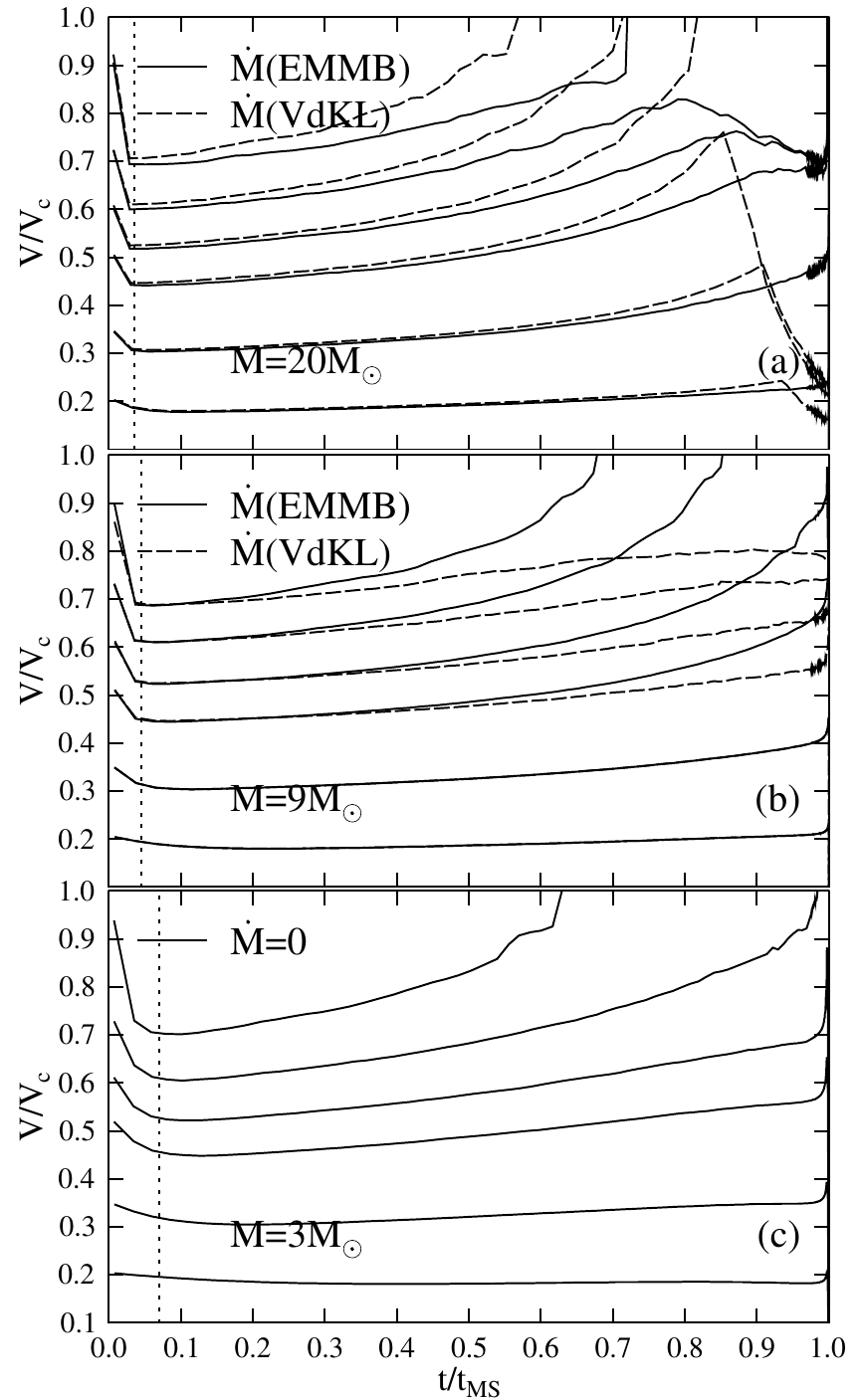}} 
\caption{\label{vvc_th} Theoretical evolution of velocity ratios $V/V_{\rm crit}$ in the MS calculated by \citet{Ekstrom2008} with mass-loss rates $\dot{M}(EMMB)$ (full lines) and $\dot{M}(VdKL)$ (dashed lines) for $M=3$, 9, and $20M_{\odot}$. The vertical dotted lines near the ZAMS indicate the fractional age $t/t_{\rm MS}$ to which correspond the adopted ZAMS velocity ratios $V_{\rm ZAMS}/V_{\rm c}$.}
\end{figure}

\subsection{Inference of $V/V_{\rm c}$ velocity ratios in the ZAMS }\label{izamsvvc} 

\begin{figure*}[] 
\centerline{\includegraphics[scale=0.8]{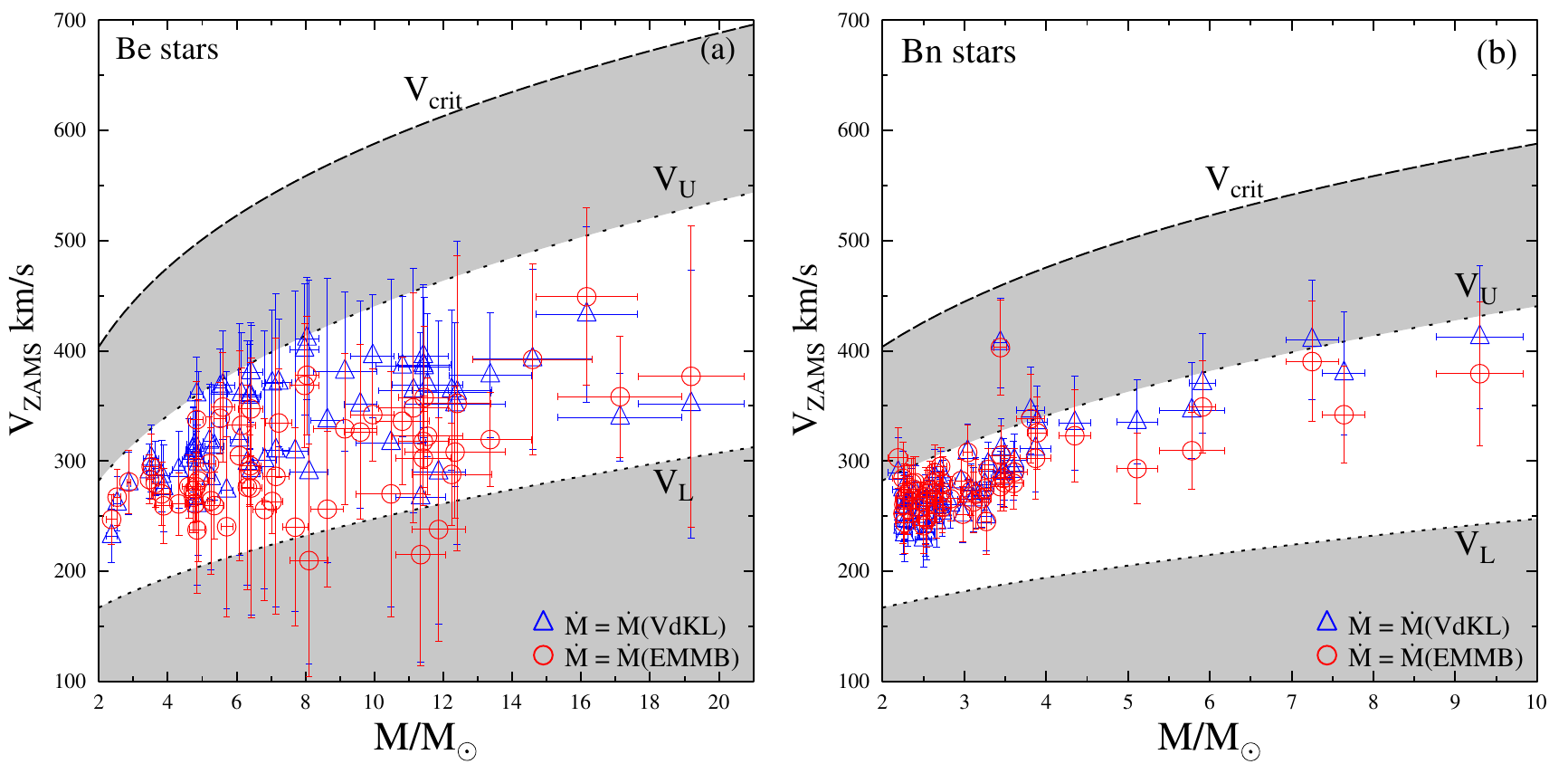}}
\caption{\label{vzams_v} Equatorial rotational velocities $V_{\rm ZAMS}$ of Be [panel (a)] and Bn [panel (b)] stars deduced using models of stellar evolution by \citet{Ekstrom2008} with mass-loss rates $\dot{M}(EMMB)$ (red circles) and $\dot{M}(VdKL)$ (blue triangles). In the figure is shown the curve $V_{\rm crit}$ in the ZAMS as a function of the stellar mass and the limiting curves $V_{\rm L}$ and $V_{\rm U}$ that surround the deduced $V_{\rm ZAMS}$ values.} 
\end{figure*}

Model predictions of changes of the surface rotational velocity strongly depend on the prescribed mass-loss rate \citep{Ekstrom2012}.
%These authors find that number of prescriptions for the mass-loss rate exist, which differentiate from each other such as the stellar mass and evolutionary phase.
Mass loss is indeed a key phenomenon that controls the evolution of stars, so that high uncertainties may also affect the final results accordingly. For an initial look at the consequences from the evolution of rotational velocities as a consquence of the mass-loss phenomenon, it is interesting to use different prescriptions for the mass-loss rate for the whole sample of stars studied in this work (Be and Bn stars). \citet{Ekstrom2008} published two series of models for stars evolving with several initial rotational velocities and chemical compositions for the metallicity $Z=0.02$. The first series of models depend on the mass-loss rates given by \citet{deJager1988} and \citet{Kudritzki2000}, which in this work we call  EMMB \citep{Ekstrom2008}; these rates are specified hereafter as $\dot{M}(EMMB)$ mass-loss rates. The second series of models rely on the mass-loss rates suggested by \citet{Vink2000}, which we call VdKL [VdKL=\citep{Vink2000}], or $\dot{M}(VdKL)$. These two mass-loss prescriptions differ somewhat at epoch $t/t_{\rm MS} \gtrsim 0.5$ and carry sensitive differences in the evolution of the $V/V_{\rm crit}$ velocity ratios by the end of the MS phase. These effects can be seen in Fig.~\ref{vvc_th} and affect stars with masses $M \gtrsim 9~M_{\odot}$. \par

\begin{figure*}[]
\centerline{\includegraphics[scale=0.85]{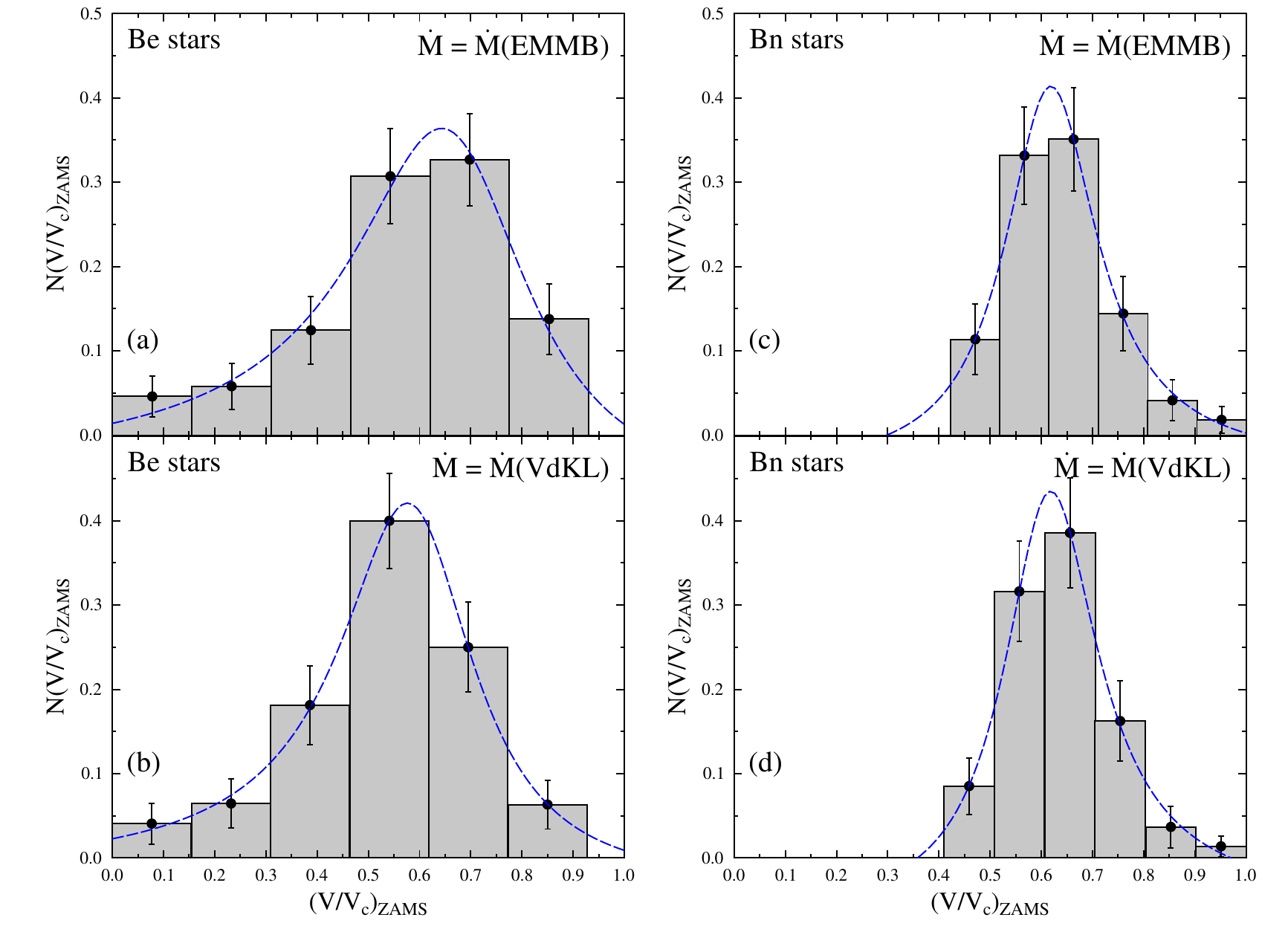}}
\caption{\label{vzams_vvc_stat} Normalized histograms showing the distribution of ratios $V/V_{\rm c}$ in the ZAMS derived for the program Be [panels (a) and (b)] and Bn [panels (c) and (d)] stars. The results shown in panels (a) and (c) were obtained with EMMB mass-loss rates, while in panels (b) and (d) are for VdKL mass-loss rates. Error bars indicate sampling uncertainties according to the errors associated with the measured $V\!\sin i$ parameters. Fitted Person distributions are superimposed (blue dashed curves) to better perceive the differences in the asymmetries of distributions for each type of stars.}
\end{figure*}

In each series of models we can adopt two different values to represent $V_{\rm ZAMS}$. There is the absolute initial $V_{\rm ZAMS}$ rotational velocity assigned to the models at the nominal $t/t_{\rm MS} = 0$. There is also the value attained by the star once the first phase of angular momentum redistribution occurs, which enables the object to acquire a stabilized rotational law. These phases last from 1\% to 2\% of the whole MS period, $t/t_{\rm MS}\sim0.01-0.02$ \citep{Maeder2000}. We adopted the second value, because in real stars such a stabilization takes place in the pre-MS evolution, far before the stars cross the ZAMS to enter the long quasi-stationary MS phase. \par

The iteration of the system of equations Eq.~\eqref{eq_8} with the use of the relations shown in Fig.~\ref{vvc_th} enables us to determine $V_{\rm ZAMS}$ of our program Be and Bn stars as a function of both mass-loss rates prescriptions, $\dot{M}(EMMB)$ and $\dot{M}(VdKL)$. The results obtained are presented in Fig.~\ref{vzams_v}a for Be stars and Fig.~\ref{vzams_v}b for Bn stars. We see that there is a tendency for a shift of points dependent on  $\dot{M}(EMMB)$ toward slightly lower $V_{\rm ZAMS}$ values than those calculated using models with $\dot{M}(VdKL)$ mass-loss rates for stellar masses $M \lesssim 12M_{\odot}$. The effect seems to be more marked among Be stars. The effect is less significant for Bn stars simply because the mass-loss rates are lower or zero for objects with masses $M \lesssim 4M_{\odot}$ that concern most of our program Bn stars. Common average lower and upper limiting curves, $V_{\rm L}$ and $V_{\rm U}$, respectively, are represented in Figs. 18a and b. These limiting curves were determined by searching to include the highest possible number of stars inside the region. It is however interesting to note that although both type of objects begin their MS evolutionary phase with similar rotational velocities, only a fraction of these objects will at some moment display the Be phenomenon. \par

Another way of detecting possible differences in the ZAMS rotational velocity between Be and Bn stars and the effects carried by different prescriptions of mass-loss rates is shown in Fig.~\ref{vzams_vvc_stat}, where the histograms of ratios $(V/V_{\rm c})_{\rm ZAMS}$ are indicated. It is seen in this figure that for Be stars the values of $(V/V_{\rm c})_{\rm ZAMS}$ are slightly lower when the $\dot{M}(VdKL)$ mass-loss rates are used, while differences are hardly noticeable for Bn stars. This is because only a few, the more massive Bn stars,  undergo mass-loss phenomena. We also note that the skewness of the Be star distributions is negative, similar to what can be noted in Fig.~\ref{vvc_hoy} for the distribution of $V/V_{\rm c}$ corresponding to the current stellar evolutionary stage. Contrarily, the skewness of the Bn star distributions becomes positive (mean and median are larger than the mode), while it is likely negative in the respective distribution of $V/V_{\rm c}$ corresponding to the current stellar evolutionary stage shown in Fig.~\ref{vvc_hoy}b. \par

We treat Eq.~\eqref{eq_8} by assuming that the input parameters are independent, therefore solutions of the $\Omega/\Omega_{\rm c}$ ratio are obtained for each star. In the case in which $\Omega/\Omega_{\rm c}=0.95$ for all stars, the global aspect of the distribution of points is similar to Fig.~\ref{vzams_v} apart from punctual positions of points, and we do not show this. This is also true for the asymmetries of the distributions shown in Fig.~\ref{vzams_vvc_stat}. Owing to the number of uncertainties that mar our statistics, the results obtained must be taken just as indications for possibly differences or similarities between the Be and Bn star populations. Since they indicate that the distributions of rotational velocities could not be the same in the ZAMS, it would be interesting to carry out this type of study again with a higher number of objects, where all the parameters entering equations Eq.~\eqref{eq_8} could be considered entirely independent. This would then enable us to establish the actual characteristic of distributions. If differences in their aspects could be confirmed, detailed studies carried out on the redistribution of internal angular momentum based on the derived initial distribution of rotational velocities would provide information to answer the question as to why some B-type rapid rotators in ZAMS cannot become Be. \par

\section{Conclusions}

In this paper we present observational characteristics of a sample of Be and Bn stars based on the behavior of their sBD, the H$\alpha$ emission intensity, and their correlation with the stellar parameters (Sects.~\ref{obsr} and \ref{sbdi}). Our H$\alpha$ observations of the program Bn stars revealed that six stars previously considered in the literature as Bn stars, \object{HD\,31209}, \object{HD\,42327}, \object{HD\,43445}, \object{HD\,165910}, \object{HD\,171623,} and \object{HD\,225132}, are  genuine Be stars. Since the rapid rotation characterizes both Be and Bn stars, the occasional appearance of some emission in the H$\alpha$ Balmer line of Bn stars and the presence of signatures such as a sBD were motivating reasons to think that Bn and Be probably belong to the same class of objects. \par

We obtained low-resolution spectra in the visual region of the program stars and determined their ($\lambda_1, D^*$) parameters as defined in the BCD spectrophotometric stellar classification system. With these quantities, but also with the help of other methods and calibrations of stellar fundamental parameters, we determined the apparent fundamental parameters ($T_{\rm eff},\log L/L_{\odot}$). Furthermore we measured the $V\!\sin i$ for a number of stars. Using the models of stellar evolution we also obtained their masses, radii, surface gravities, ages, and critical equatorial velocities ($M/M_{\odot},R/R_{\odot},\log g,t/t_{\rm MS},V_{\rm c}$) (Sect.~\ref{apppar}). \par

The correlations between the emission quantities and the apparent fundamental parameters gave us quantified tendencies, but these do not reveal tight correlations. Limits regarding effective temperatures and rotation velocities were found to characterize the appearance of the sBD in emission or in absorption (Sects.~\ref{apppar} and \ref{rbeiafp}). A slightly more detailed statistical treatment of the apparent $V\!\sin i$ parameters showed that the sBD in emission appear at angles lower than some 70\degr, while they appear in absorption at angles larger than some 60\degr\,(Sect.~\ref{sbdi}). This suggests that the CE layers close to the central star should have a large enough vertical height. In addition, we found than Be stars without a sBD generally present small intensities in the H$\alpha$ line and that the intensity of the H$\alpha$ line become stronger as $T_{\rm eff}$ increases when the sBD is in emission. \par

Because all the apparent fundamental parameters are not entirely independent, we estimated an order of magnitude of the correction we had to introduce for rotationally induced effects by assuming that all stars rotate at the angular velocity rate $\Omega/\Omega_{\rm c}=0.95$ (Sect.~\ref{pnrcpar}). We could then place our program Be and Bn stars on an HR diagram and study their evolutionary status. We conclude that the main difference between Be and Bn stars is that Bn stars have masses $M \lesssim 9M_{\odot}$, while Be stars range from $3M_{\odot} \lesssim M $ to $M \lesssim 20M_{\odot}$. We also note that Bn stars of masses $4 \lesssim M/M_{\odot} \lesssim 10$ are close to the ZAMS, while in this mass interval Be stars occupy the entire MS evolutionary span. Bn stars are more numerous than Be stars for masses $M/M_{\odot} \lesssim 4$, but both roughly share the same evolutionary domain (Sect.~\ref{hrbebn}). From the study of the distribution of stellar ages we conclude that Bn stars can have age ratios $0.2 \lesssim (t/t_{\rm MS})_{\rm Bn} \lesssim 0.8$ if $M/M_{\odot} \lesssim 4$, whereas they have $0.0 \lesssim (t/t_{\rm MS})_{\rm Bn} \lesssim 0.2$ for masses are $M/M_{\odot} \gtrsim 5$ (Sect.~\ref{dabebn}). Thus, it should not be excluded that the most massive Bn stars could at any time display the Be phenomenon. \par

To characterize Be and Bn stars as fast rotators and thus find possible stellar population similarities or differences, we studied the distributions of true rotational velocity ratios $V/V_{\rm c}$ of our program stars at the current stellar evolutionary stage and in the ZAMS. We noted, in particular, biases that may be affecting the statistics due to selection effects concerning the aspect angles of the program stars. In spite of these difficulties, which we cannot overcome at the moment, initial indications seem to exist for differences in the velocity distributions. At the current stellar evolutionary stage, they have roughly the same global aspect (modes and negative skewness, as shown in Sect.~\ref{davbebn}), but their respective skewnesses do not have the same sign in the ZAMS; this sign is negative for Be stars and positive for Bn stars (Sect.~\ref{zamsdv}). Studies similar to this work, but with more detail regarding the fundamental parameters and including a larger number of objects, might confirm our findings and succeed in deciding whether or not both type of objects are of the same class. \par

Finally, we comment on the disk structure near the central star that is suggested by some results of the present paper. Detailed accounts of successful representations of the observed emission in the H$\alpha$ line of Be stars and of the continuum energy distribution from the far-UV to the IR, as well as of the polarization in the visible spectral range, are given in a series of rather recent papers and references therein \citep{Araya2017,Arcos2017,Klement2017,Marr2018}. In all these studies, the physical structure of the CE of Be stars is studied in the frame of an axisymmetric quasi-Keplerian disk in hydrostatic equilibrium in the vertical axis, where the scale height $H$ is proportional to $R^{3/2}T^{1/2}$ and $R$ represents the equatorial radius of the disk and $T$ is its local temperature. However, from statistical arguments based on the study of the distribution of the $V\!\sin i$ parameter, we conclude in the present contribution that the sBD in emission could be seen in Be stars at inclination angles down to $i\sim70\degr$, while this component should be in absorption in Be stars that have inclination angles $i \gtrsim 60\degr$. This finding may then conflict with the geometrically thin vertical structure of the density distribution of the above-mentioned flaring disks near the central star, where the sBD is raised. Thus, we could suggest an enlarged vertical scale height of the circumstellar disk near the central star. This idea is supported by the following results by different authors. \par

Simple models of circumstellar disks with electron densities above $N_{\rm e}=10^{13}$ cm$^{-3}$ near the central star can explain the increase of a sBD. However, at such gas densities, the sBD strongly encroaches the photospheric BD, which is rarely observed. To overcome the need for gas densities of disks near the star that are too high, a larger vertical scale of the disk than usually foreseen for pure Keplerian disks could then be envisioned. On the other hand, owing to the very rapid drop of the non-LTE source function of \ion{Fe}{ii} lines, two orders of magnitude within two stellar radii, \citet{Arias2006} and \citet{Zorec2007_470},  also suggested enlarged vertical scale heights. These zones are roughly the same where the sBD is formed. Also, larger vertical density scale heights can exist if the presence of possible disordered magnetic fields of low intensity is not neglected and if sources of heating energy in the disk near the central star are taken into account. In a series of papers, \citet[][and references therein]{Kurf2018} show that an equilibrium disk structure with enhanced density and a temperature increase up to some $T\sim10^5$ K within 2 stellar radii is obtained through energy dissipation by viscosity. Such temperatures are able to increase the vertical scale heights of disks. Unfortunately, up to now there is neither observational evidence for such magnetic fields nor thorough hydro-dynamical calculations to confirm such dissipations in disks with enhanced vertical scale heights. Finally, we can ask whether smooth Keplerian density structures of disks near the star can survive when different types of mass ejections in Be stars takes place. In these objects continuous and variable winds with average rates \mbox{$\dot{M}\sim10^{-9}$ $M_{\odot}$/yr} (Snow 1982) can coexist together with discrete mass ejections of $\Delta M\lesssim10^{-10}$~$M_{\odot}$ \citep{Guinan1984, Brown1992, Hanuschik1993, Floquet2000, Hubert2000, Zorecetal2000}. In such circumstances, the circumstellar environment may look very clumpy. The interaction of winds with the circumstellar clumps can build particular radial density structures \citep{Arthur1994, Meilland2006} and produce shock fronts where $T_{\rm e}\gg T_{\rm eff}$ \citep{Hartquist1986, Dyson1992, Arthur1994}. \par

These suggestions are speculations that may invite debate as well as requiring observational support and detailed hydrodynamical calculations to be accepted or rejected. It would then be advisable that in the future we proceed to model circumstellar disks with more physical requirements and try to model several spectral lines of different elements (\ion{Fe}{ii} emission lines, H$\alpha$, H$\beta$, H$\gamma$ to represent Balmer decrements, together with other IR emission lines) observed simultaneously, and other signatures of the disk, such as the sBD, which are sensitive to the up to now widely neglected physical inputs in circumstellar disks. Questions related to disk-wind interactions perhaps require observations made in the far-UV radiation, which at this time may seem unlikely. \par

\begin{acknowledgements}
We would like to thank Carol Jones, our referee, for her valuable comments and suggestions, which helped to improve our manuscript.\\
This research has made use of the SIMBAD database, operated at CDS, Strasbourg, France. This work has made use of the BeSS database, operated at LESIA, Observatoire de Meudon, France: http://basebe.obspm.fr. We especially thank the following observers who have submitted data to BeSS that we have used in this publication: Christian Buil, Terry Bohlsen, André Favaro, Joan Guarro Fló, Bernard Heathcote and Nando Romeo.
We thank also Christian Buil for his reduced individual H$\alpha$ observations available via The spectroscopic Be-stars Atlas, http://www.astrosurf.com/buil/us/becat.htm.\\
L.C. acknowledges financial support from CONICET (PIP 0177), the Agencia Nacional de Promoción Científica y Tecnológica (Préstamo BID PICT 2016-1971) and the Programa de Incentivos (G11/137) of the Universidad Nacional de La Plata (UNLP), Argentina. A.G. acknowledges the financial support received from the Agencia Nacional de Promoción Científica y Tecnológica of Argentina (PICT2017-3790).\\

\end{acknowledgements}

\bibliographystyle{../../paquetes/aa-package/bibtex/aa}

\bibliography{sbd.bib}

% \begin{appendix} 

\onecolumn

\begin{longtable}{llllll}
\caption{Observed Be stars. All the spectra were obtained in CASLEO, with the exception of those denoted with $a$ \citep[BeSS;][]{Neiner2011} or $b$ \citep{Atlas}.}\label{tabla_Be}\\
\hline\hline
\noalign{\smallskip}
HD  &  \multicolumn{1}{c}{$\alpha$} & \multicolumn{1}{c}{$\delta$} & \multicolumn{1}{c}{V} & \multicolumn{2}{c}{Observation dates} \\ 
    &  \multicolumn{1}{c}{(ICRS,J2000.0)} & \multicolumn{1}{c}{(ICRS,J2000.0)}           & [mag]    & Low-resolution & High-resolution H$\alpha$ \\
\hline
\noalign{\smallskip} 
\endfirsthead
\caption{Continued.} \\
\noalign{\smallskip} 
\hline
\noalign{\smallskip}
HD  &  \multicolumn{1}{c}{$\alpha$} & \multicolumn{1}{c}{$\delta$} & \multicolumn{1}{c}{V} & \multicolumn{2}{c}{Observation dates} \\ 
    &  \multicolumn{1}{c}{(ICRS,J2000.0)} & \multicolumn{1}{c}{(ICRS,J2000.0)}           & [mag]    & Low-resolution & High-resolution      \\
\noalign{\smallskip}
\hline
\noalign{\smallskip}
\endhead
\noalign{\smallskip}
\endfoot
\object{HD 10144} & 01:37:42.84548 & $-$57:14:12.3101 & 0.46 & 03-09-04 & 03-09-10$^{a}$\\
\object{HD 16582} & 02:39:28.95579 & +00:19:42.6345 & 4.07 & 31-08-04 & --\\
\object{HD 23016} & 03:42:18.94710 & +19:42:00.9206 & 5.69 & 31-08-04 & 07-09-02$^{a}$\\
\object{HD 28248} & 04:24:06.90247 & $-$57:15:11.3797 & 7.52 & 31-08-04 & --\\
\object{HD 28497} & 04:29:06.92534 & $-$13:02:54.1234 & 5.41 & 31-08-04/11-02-14 & 11-12-06$^{b}$/10-02-14\\
\object{HD 30076} & 04:44:05.32090 & $-$08:30:12.8425 & 5.81 & 03-09-04/11-02-14 & 14-12-03$^{b}$/10-02-14\\
\object{HD 30739} & 04:50:36.72298 & +08:54:00.6493 & 4.35 & 02-09-04 & 11-12-16$^{a}$\\
\object{HD 35165} & 05:21:16.86051 & $-$34:20:42.1863 & 6.09 & 11-02-14 & --\\
\hline\hline 
\end{longtable}

\begin{longtable}{lllllc}
\caption{Observed Bn stars. All the spectra were obtained in CASLEO.} \label{tabla_Bn}\\
\hline\hline
\noalign{\smallskip}
HD  &  \multicolumn{1}{c}{$\alpha$} & \multicolumn{1}{c}{$\delta$} & \multicolumn{1}{c}{V} & \multicolumn{2}{c}{Observation dates} \\ 
    &  \multicolumn{1}{c}{(ICRS,J2000.0)} & \multicolumn{1}{c}{(ICRS,J2000.0)}           & [mag]    & Low-resolution & High-resolution      \\
\hline
\noalign{\smallskip} 
\endfirsthead
\caption{Continued.} \\
\noalign{\smallskip} 
\hline
\noalign{\smallskip}
HD  &  \multicolumn{1}{c}{$\alpha$} & \multicolumn{1}{c}{$\delta$} & \multicolumn{1}{c}{V} & \multicolumn{2}{c}{Observation dates} \\ 
    &  \multicolumn{1}{c}{(ICRS,J2000.0)} & \multicolumn{1}{c}{(ICRS,J2000.0)}           & [mag]    & Low-resolution & High-resolution      \\
\noalign{\smallskip}
\hline
\noalign{\smallskip}
\endhead
\noalign{\smallskip}
\endfoot
\object{HD 560} & 00:10:02.20293 & +11:08:44.9280 & 5.53 & 31-08-04 & --\\
\object{HD 5617} & 00:57:37.76060 & $-$18:59:55.0309 & 6.9 & 31-08-04 & --\\
\object{HD 10161} & 01:38:49.99449 & $-$25:01:19.9155 & 6.69 & 31-08-04 & --\\
\object{HD 10894} & 01:47:09.10651 & +10:50:39.2638 & 7.05 & 31-08-04 & --\\
\object{HD 15130} & 02:25:57.00560 & $-$12:17:25.7104 & 4.87 & 31-08-04 & --\\
\object{HD 18331} & 02:56:37.42306 & $-$03:42:44.3505 & 5.16 & 01-09-04 & --\\
\object{HD 18546} & 02:57:32.62737 & $-$38:11:27.3204 & 6.41 & 01-09-04 & --\\
\object{HD 21364} & 03:27:10.15071 & +09:43:57.6343 & 3.75 & 01-09-04 & --\\
\hline\hline 
\end{longtable}

\setlength{\tabcolsep}{3pt} 
\begin{longtable}{lcclccrcrrccrrc}
\caption{Observed BCD and apparent parameters of Be stars.} \label{tabla1_Be}\\
\noalign{\smallskip}
\hline\hline
\noalign{\smallskip}
HD & D$_*$ & $\lambda_1$ & SpTLC & $d$ & $E(B-V)$ & M$_{\rm V}$ & $\sigma_{M_{\rm V}}$ & T$_{\rm eff}$ & 
$\sigma_{T_{\rm eff}}$ &
$\log L/L_{\odot}$ & $\sigma_{\log L/L_{\odot}}$ & $V\!\sin i$ & $\sigma_{V\!\sin i}$ & Ref. \\ 
   & [dex] & [\AA]       &       &[dex]& [mag]  & [mag]       & [mag]                &  [K]      &  
[K]                & [dex]                       &  [dex]      &  [km/s]                & [km/s]           &     \\
\hline
\noalign{\smallskip} 
\endfirsthead
\caption{Continued.} \\
\noalign{\smallskip} 
\hline
\noalign{\smallskip}
HD & D$_*$ & $\lambda_1$ & SpTLC & $d$ & $E(B-V)$ & M$_{\rm V}$ & $\sigma_{M_{\rm V}}$ & T$_{\rm eff}$ & 
$\sigma_{T_{\rm eff}}$ &
$\log L/L_{\odot}$ & $\sigma_{\log L/L_{\odot}}$ & $V\!\sin i$ & $\sigma_{V\!\sin i}$ & Ref. \\ 
\noalign{\smallskip}
\hline
\noalign{\smallskip}
\endhead
\noalign{\smallskip}
\endfoot
  \object{HD  10144} & 0.22 & 37 & B4-5III   &  0.03&  0.02 & -2.68 & 0.17 & 16697 & 1119 & 3.574 & 0.060 & 235 & 28 &  1 \\
  \object{HD  16582} & 0.14 & 49 & B2III     &  0.00&  0.02 & -3.11 & 0.46 & 22885 & 1888 & 4.057 & 0.129 &  12 &  1 &  2 \\
  \object{HD  23016} & 0.38 & 48 & B8IV      &  0.00&  0.10 & -0.52 & 0.22 & 12111 &  257 & 2.419 & 0.057 & 282 & 33 &  3 \\
  \object{HD  28248} & 0.13 & 49 & B2III     &  0.00&  0.18 & -3.46 & 0.15 & 20000 & 2000 & 3.390 & 0.055 &  30 &  3 &  3 \\
  \object{HD  28497} & 0.09 & 65 & B0.5V     & -0.03/-0.06&  0.04 & -3.42 & 0.39 & 30588 & 1912 & 4.484 & 0.108 & 335 & 40 &  1 \\
  \object{HD  30076} & 0.11 & 37 & B2II      & -0.07/0.00&  0.09 & -5.24 & 0.37 & 21278 & 2202 & 4.837 & 0.114 & 225 & 27 &  4 \\
  \object{HD  30739} & 0.54 & 64 & A1-2IV    &  0.00&  0.01 &  0.71 & 0.56 &  8781 &  349 & 1.700 & 0.143 & 212 & 25 &  3 \\
  \object{HD  31209} & 0.53 & 58 & A1IV      &  0.02/0.03&  0.04 &  0.46 & 0.44 &  9204 &  251 & 1.824 & 0.112 & 270 & 32 &  3 \\
\hline
\end{longtable}

\setlength{\tabcolsep}{3pt}
\begin{longtable}{lcclccrcrrccrrc}
\caption{Observed BCD and apparent parameters of Bn stars.} \label{tabla1_Bn}\\
\noalign{\smallskip}
\hline\hline
\noalign{\smallskip}
HD & D$_*$ & $\lambda_1$ & SpTLC & $d$ & $E(B-V)$ & M$_{\rm V}$ & $\sigma_{M_{\rm V}}$ & T$_{\rm eff}$ & 
$\sigma_{T_{\rm eff}}$ &
$\log L/L_{\odot}$ & $\sigma_{\log L/L_{\odot}}$ & $V\!\sin i$ & $\sigma_{V\!\sin i}$ & Ref. \\ 
   & [dex] & [\AA]       &       &[dex]& [mag]  & [mag]       & [mag]                &  [K]      &  
[K]                & [dex]                       &  [dex]      &  [km/s]                & [km/s]           &     \\
\hline
\noalign{\smallskip} 
\endfirsthead
\caption{Continued.} \\
\noalign{\smallskip} 
\hline
\noalign{\smallskip}
HD & D$_*$ & $\lambda_1$ & SpTLC & $d$ & $E(B-V)$ & M$_{\rm V}$ & $\sigma_{M_{\rm V}}$ & T$_{\rm eff}$ & 
$\sigma_{T_{\rm eff}}$ &
$\log L/L_{\odot}$ & $\sigma_{\log L/L_{\odot}}$ & $V\!\sin i$ & $\sigma_{V\!\sin i}$ & Ref. \\ 
\noalign{\smallskip}
\hline
\noalign{\smallskip}
\endhead
\noalign{\smallskip}
\endfoot
     \object{HD 560} & 0.44 & 63 & B9V       &  0.00 & 0.03 &  0.62 & 0.22 & 10666 &  291 & 1.862 & 0.057 & 249 & 29 &  3 \\
    \object{HD 5617} & 0.50 & 76 & A2V       &  0.00 & 0.04 &  0.97 & 0.51 &  9151 &  369 & 1.619 & 0.130 & 236 & 28 &  1 \\
   \object{HD 10161} & 0.43 & 47 & B9IV      &  0.00 & 0.00 & $-$0.23 & 0.06 & 11187 &  244 & 2.237 & 0.019 & 160 & 19 &  3 \\
   \object{HD 10894} & 0.50 & 54 & A0IV      &  0.00 & 0.01 &  0.21 & 0.38 &  9861 &  285 & 1.970 & 0.096 & 212 & 30 &  1 \\
   \object{HD 15130} & 0.52 & 51 & A0-1IV    &  0.00 & 0.01 &  0.36 & 0.29 &  9579 &  209 & 1.889 & 0.074 & 219 & 26 &  3 \\
   \object{HD 18331} & 0.49 & 65 & A1V       &  0.00 & 0.04 &  1.05 & 0.37 &  9666 &  367 & 1.620 & 0.095 & 236 & 28 &  3 \\
   \object{HD 18546} & 0.49 & 65 & A0-1V     &  0.00 & 0.00 &  0.93 & 0.29 &  9666 &  336 & 1.667 & 0.075 & --- & -- &  1 \\
   \object{HD 21364} & 0.35 & 59 & B7-8V     &  0.00 & 0.04 & $-$0.11 & 0.26 & 12430 &  475 & 2.274 & 0.069 & 224 & 26 &  3 \\
\hline
\end{longtable}

\setlength{\tabcolsep}{5pt}
\begin{longtable}{lrrccrrcccc}
\caption{Derived apparent fundamental parameters of Be stars} \label{tabla2_Be}\\
\noalign{\smallskip}
\hline\hline
\noalign{\smallskip}
HD & $M/M_{\odot}$ & $\sigma_{M/M_{\odot}}$ & $\log g$ & $\sigma_{\log g}$ & $R/R_{\odot}$ & $\sigma_{R/R_{\odot}}$ &
$V_{\rm c}$ & $\sigma_{V_{\rm c}}$ & $t/t_{\rm MS}$ & $\sigma_{t/t_{\rm MS}}$ \\
   &               &                        & [dex]    & [dex]             &               &                        & 
 [km/s]       &  [km/s]                &                &                         \\
\hline
\noalign{\smallskip} 
\endfirsthead
\caption{Continued.} \\
\noalign{\smallskip} 
\hline
\noalign{\smallskip}
HD & $M/M_{\odot}$ & $\sigma_{M/M_{\odot}}$ & $\log g$ & $\sigma_{\log g}$ & $R/R_{\odot}$ & $\sigma_{R/R_{\odot}}$ &
$V_{\rm c}$ & $\sigma_{V_{\rm c}}$ & $t/t_{\rm MS}$ & $\sigma_{t/t_{\rm MS}}$ \\
\noalign{\smallskip}
\hline
\noalign{\smallskip}
\endhead
\noalign{\smallskip}
  \object{HD  10144} &  6.94 & 0.18 & 3.55 & 0.09 &  7.32 &  1.96 & 350 & 51 & 1.019 & 0.006 \\
  \object{HD  16582} & 10.37 & 0.61 & 3.79 & 0.13 &  6.79 &  2.24 & 445 & 86 & 0.886 & 0.114 \\
  \object{HD  23016} &  3.66 & 0.07 & 3.87 & 0.04 &  3.68 &  0.31 & 359 & 18 & 0.880 & 0.054 \\
  \object{HD  28248} &  7.14 & 0.35 & 4.06 & 0.14 &  4.13 &  1.65 & 476 & 71 & 0.581 & 0.240 \\
  \object{HD  28497} & 16.06 & 0.89 & 4.05 & 0.11 &  6.22 &  1.55 & 584 & 89 & 0.510 & 0.193 \\
  \object{HD  30076} & 16.84 & 1.04 & 3.09 & 0.13 & 19.29 &  7.98 & 335 & 79 & 1.019 & 0.001 \\
  \object{HD  30739} &  2.37 & 0.09 & 3.84 & 0.10 &  3.06 &  0.50 & 316 & 31 & 0.933 & 0.092 \\
  \object{HD  31209} &  2.55 & 0.08 & 3.83 & 0.07 &  3.21 &  0.36 & 320 & 22 & 0.940 & 0.078 \\
\hline
\end{longtable}

\setlength{\tabcolsep}{3pt}
\begin{longtable}{lrrccrrcccc}
\caption{Derived apparent fundamental parameters of Bn stars} \label{tabla2_Bn}\\
\noalign{\smallskip}
\hline\hline
\noalign{\smallskip}
HD & $M/M_{\odot}$ & $\sigma_{M/M_{\odot}}$ & $\log g$ & $\sigma_{\log g}$ & $R/R_{\odot}$ & $\sigma_{R/R_{\odot}}$ &
$V_{\rm c}$ & $\sigma_{V_{\rm c}}$ & $t/t_{\rm MS}$ & $\sigma_{t/t_{\rm MS}}$ \\
   &               &                        & [dex]    & [dex]             &               &                        & 
 [km/s]       &  [km/s]                &                &                         \\
\hline
\noalign{\smallskip} 
\endfirsthead
\caption{Continued.} \\
\noalign{\smallskip} 
\hline
\noalign{\smallskip}
HD & $M/M_{\odot}$ & $\sigma_{M/M_{\odot}}$ & $\log g$ & $\sigma_{\log g}$ & $R/R_{\odot}$ & $\sigma_{R/R_{\odot}}$ &
$V_{\rm c}$ & $\sigma_{V_{\rm c}}$ & $t/t_{\rm MS}$ & $\sigma_{t/t_{\rm MS}}$ \\
\noalign{\smallskip}
\hline
\noalign{\smallskip}
\endhead
\noalign{\smallskip}
  \object{HD    560} &  2.74 & 0.05 & 4.08 & 0.05 &  2.50 &  0.27 & 378 & 24 & 0.647 & 0.068 \\
  \object{HD   5617} &  2.34 & 0.09 & 3.99 & 0.09 &  2.57 &  0.42 & 344 & 34 & 0.777 & 0.105 \\
  \object{HD  10161} &  3.30 & 0.04 & 3.87 & 0.03 &  3.50 &  0.31 & 349 & 17 & 0.881 & 0.042 \\
  \object{HD  10894} &  2.79 & 0.07 & 3.84 & 0.07 &  3.31 &  0.39 & 330 & 23 & 0.920 & 0.077 \\
  \object{HD  15130} &  2.67 & 0.06 & 3.86 & 0.05 &  3.20 &  0.28 & 329 & 18 & 0.912 & 0.068 \\
  \object{HD  18331} &  2.38 & 0.07 & 4.09 & 0.07 &  2.30 &  0.35 & 367 & 33 & 0.642 & 0.105 \\
  \object{HD  18546} &  2.43 & 0.06 & 4.05 & 0.06 &  2.43 &  0.34 & 361 & 29 & 0.696 & 0.078 \\
  \object{HD  21364} &  3.47 & 0.08 & 4.04 & 0.06 &  2.96 &  0.45 & 391 & 34 & 0.691 & 0.084 \\
\hline
\end{longtable}

\begin{longtable}{lrrrr}
\caption{H$\alpha$ line emission equivalent widths and fluxes of Be stars} \label{tabla_flux}\\
\hline\hline
\noalign{\smallskip}
HD  & \multicolumn{1}{c}{$W_{\rm e}$} & \multicolumn{1}{c}{$W$}  & \multicolumn{1}{c}{F$_{\rm H\alpha}/10^9$} & \multicolumn{1}{c}{d}  \\
    & \multicolumn{1}{c}{[\AA]} & \multicolumn{1}{c}{[\AA]}       & \multicolumn{1}{c}{[cgs]}             & \multicolumn{1}{c}{[dex]} \\
\hline
\noalign{\smallskip}
\endfirsthead
\caption{Continued.} \\
\noalign{\smallskip}
\hline
\noalign{\smallskip}
HD  & \multicolumn{1}{c}{$W_{\rm e}$} & \multicolumn{1}{c}{$W$}  & \multicolumn{1}{c}{F$_{\rm H\alpha}/10^9$} & \multicolumn{1}{c}{d}  \\
\noalign{\smallskip}
\hline
\noalign{\smallskip}
\endhead
\noalign{\smallskip}
\endfoot    
\object{HD 10144}  &  $-$2.83  &   1.50  &  0.033  &  0.03  \\
\object{HD 23016}  &  $-$4.53  &   4.44  &  0.068  &  0.00  \\
\object{HD 28497}  &  7.42   &  10.07  &  0.692  &  $-$0.06   \\
\object{HD 28497}  &  14.25  &  16.90  &  1.161  &  $-$0.03   \\
\object{HD 30076}  &  6.62   &   9.01  &  0.299  &  0.00  \\
\object{HD 30076}  &  37.21  &  39.60  &  1.312  &  $-$0.07   \\
\object{HD 30739}  &  $-$8.22  &   5.04  &  0.035  &  0.00  \\
\object{HD 31209}  &  $-$3.11  &  $-$0.65  &  $-$0.005  &  0.03  \\
\hline
\end{longtable}

\setlength{\tabcolsep}{3pt}
\begin{longtable}{lrrccccrccccccc}
\caption{Parent nonrotating counterpart (pnrc) fundamental parameters of Be stars} \label{tabla3_Be}\\
\noalign{\smallskip}
\hline\hline
\noalign{\smallskip}
HD & T$_{\rm eff}$ & $\sigma_{T_{\rm eff}}$ & $\log L/L_{\odot}$ & $\sigma_{\log L/L_{\odot}}$ & $\log g$ & $\sigma_{\log g}$ & 
$M/M_{\odot}$ & $\sigma_{M/M_{\odot}}$ & $V\!\sin i$ & $\sigma_{V\!\sin i}$ & $V_{\rm c}$ & $\sigma_{V_{\rm c}}$ & 
$t/t_{\rm MS}$ & $\sigma_{t/t_{\rm MS}}$ \\
   &    [K]    &  [K]                   &  [dex]             &  [dex]                      &    [dex]  &  [dex]           & 
               &                        &      [km/s]   &     [km/s]             &   [km/s]       &   [km/s]    &   &   \\
\hline
\noalign{\smallskip} 
\endfirsthead
\caption{Continued.} \\
\noalign{\smallskip} 
\hline
\noalign{\smallskip}
HD & T$_{\rm eff}$ & $\sigma_{T_{\rm eff}}$ & $\log L/L_{\odot}$ & $\sigma_{\log L/L_{\odot}}$ & $\log g$ & $\sigma_{\log g}$ & 
$M/M_{\odot}$ & $\sigma_{M/M_{\odot}}$ & $V\!\sin i$ & $\sigma_{V\!\sin i}$ & $V_{\rm c}$ & $\sigma_{V_{\rm c}}$ & 
$t/t_{\rm MS}$ & $\sigma_{t/t_{\rm MS}}$ \\
\noalign{\smallskip}
\hline
\noalign{\smallskip}
\endhead
\noalign{\smallskip}
  \object{HD  10144} & 18698 & 1458 & 3.614 & 0.080 & 3.74 & 0.12 &  7.48 & 0.32 & 269 & 32 & 398 & 29 & 0.946 & 0.080 \\
  \object{HD  16582} & 23587 & 2481 & 3.869 & 0.129 & 4.00 & 0.16 &  9.78 & 0.65 &  13 &  1 & 501 & 55 & 0.632 & 0.256 \\
  \object{HD  23016} & 14066 &  472 & 2.551 & 0.077 & 4.05 & 0.06 &  4.12 & 0.10 & 320 & 38 & 411 & 14 & 0.671 & 0.088 \\
  \object{HD  28248} & 20659 & 2636 & 3.207 & 0.055 & 4.25 & 0.16 &  6.43 & 0.66 &  33 &  4 & 521 & 56 & 0.263 & 0.284 \\
  \object{HD  28497} & 32749 & 2486 & 4.406 & 0.112 & 4.24 & 0.11 & 15.75 & 0.98 & 352 & 42 & 654 & 40 & 0.162 & 0.192 \\
  \object{HD  30076} & 23915 & 3026 & 4.908 & 0.155 & 3.25 & 0.17 & 18.16 & 1.54 & 270 & 32 & 376 & 36 & 1.014 & 0.008 \\
  \object{HD  30739} &  9794 &  493 & 1.801 & 0.264 & 3.97 & 0.16 &  2.59 & 0.19 & 238 & 28 & 349 & 27 & 0.771 & 0.191 \\
  \object{HD  31209} & 10684 &  406 & 1.964 & 0.122 & 4.00 & 0.08 &  2.87 & 0.11 & 286 & 34 & 365 & 15 & 0.739 & 0.096 \\
\hline
\end{longtable}

\setlength{\tabcolsep}{3pt}
\begin{longtable}{lrrccccrccccccc}
\caption{Parent nonrotating counterpart (pnrc) fundamental parameters of Bn stars} \label{tabla3_Bn}\\
\noalign{\smallskip}
\hline\hline
\noalign{\smallskip}
HD & T$_{\rm eff}$ & $\sigma_{T_{\rm eff}}$ & $\log L/L_{\odot}$ & $\sigma_{\log L/L_{\odot}}$ & $\log g$ & $\sigma_{\log g}$ & 
$M/M_{\odot}$ & $\sigma_{M/M_{\odot}}$ & $V\!\sin i$ & $\sigma_{V\!\sin i}$ & $V_{\rm c}$ & $\sigma_{V_{\rm c}}$ & 
$t/t_{\rm MS}$ & $\sigma_{t/t_{\rm MS}}$ \\
   &    [K]    &  [K]                   &  [dex]             &  [dex]                      &    [dex]  &  [dex]           & 
               &                        &      [km/s]   &     [km/s]             &   [km/s]       &   [km/s]    &   &   \\
\hline
\noalign{\smallskip} 
\endfirsthead
\caption{Continued.} \\
\noalign{\smallskip} 
\hline
\noalign{\smallskip}
HD & T$_{\rm eff}$ & $\sigma_{T_{\rm eff}}$ & $\log L/L_{\odot}$ & $\sigma_{\log L/L_{\odot}}$ & $\log g$ & $\sigma_{\log g}$ & 
$M/M_{\odot}$ & $\sigma_{M/M_{\odot}}$ & $V\!\sin i$ & $\sigma_{V\!\sin i}$ & $V_{\rm c}$ & $\sigma_{V_{\rm c}}$ & 
$t/t_{\rm MS}$ & $\sigma_{t/t_{\rm MS}}$ \\
\noalign{\smallskip}
\hline
\noalign{\smallskip}
\endhead
\noalign{\smallskip}
  \object{HD    560} & 11719 &  427 & 1.921 & 0.069 & 4.21 & 0.06 &  2.92 & 0.06 & 258 & 31 & 416 & 15 & 0.399 & 0.128 \\
  \object{HD   5617} & 10128 &  524 & 1.701 & 0.137 & 4.11 & 0.10 &  2.51 & 0.10 & 244 & 29 & 377 & 20 & 0.590 & 0.163 \\
  \object{HD  10161} & 11825 &  332 & 2.193 & 0.026 & 4.01 & 0.04 &  3.29 & 0.03 & 170 & 20 & 379 &  8 & 0.734 & 0.046 \\
  \object{HD  10894} & 10809 &  412 & 2.023 & 0.103 & 3.98 & 0.07 &  2.96 & 0.10 & 220 & 26 & 362 & 14 & 0.770 & 0.085 \\
  \object{HD  15130} & 10560 &  327 & 1.957 & 0.084 & 3.98 & 0.06 &  2.85 & 0.08 & 226 & 27 & 361 & 11 & 0.763 & 0.065 \\
  \object{HD  18331} & 10577 &  513 & 1.672 & 0.102 & 4.22 & 0.08 &  2.51 & 0.11 & 245 & 29 & 402 & 19 & 0.398 & 0.178 \\
  \object{HD  18546} & 10586 &  456 & 1.719 & 0.075 & 4.18 & 0.07 &  2.57 & 0.06 & --- & -- & 396 & 16 & 0.475 & 0.136 \\
  \object{HD  21364} & 13445 &  661 & 2.289 & 0.076 & 4.18 & 0.08 &  3.63 & 0.10 & 241 & 28 & 431 & 21 & 0.449 & 0.158 \\
\hline
\end{longtable}

\end{document}